\documentclass[10pt,journal,compsoc]{IEEEtran}
\usepackage[compress]{cite}
\usepackage{graphicx}
\ifCLASSINFOpdf
\else
\fi
\usepackage{amsmath}
\usepackage{amssymb}

\usepackage{listings}

\ifCLASSOPTIONcompsoc
 \usepackage[caption=false,font=footnotesize,labelfont=sf,textfont=sf]{subfig}
\else
 \usepackage[caption=false,font=footnotesize]{subfig}
\fi
\usepackage{multirow}
\usepackage{makecell}
\usepackage[export]{adjustbox}

\usepackage{hyperref}
\usepackage[switch]{lineno}
\usepackage{color, colortbl}
\definecolor{Gray}{gray}{0.9}

\begin{document}
% \linenumbers
\title{Does Interacting Help Users Better Understand the Structure of Probabilistic Models?}
%~\IEEEmembership{Student Member,~IEEE,
\author{Evdoxia~Taka,
        Sebastian~Stein,
        and~John~H.~Williamson% <-this % stops a space
\IEEEcompsocitemizethanks{\IEEEcompsocthanksitem E. Taka is with the School of Computing Science, University of Glasgow, UK.\protect\\
E-mail: e.taka.1@research.gla.ac.uk
\IEEEcompsocthanksitem S. Stein and J. H. Williamson are with the School of Computing Science, University of Glasgow, UK.}% <-this % stops an unwanted space
% \thanks{Manuscript received February 19, 2022; revised May xx, xxxx.}}
\thanks{This work has been submitted to the IEEE for possible publication. Copyright may be transferred without notice, after which this version may no longer be accessible.}}


\markboth{IEEE Transactions on Visualization and Computer Graphics,~Vol.~xx, No.~x, February~2022}%
{Taka \MakeLowercase{\textit{et al.}}: Does Interaction Help Users Better Understand the Structure of Probabilistic Models?}



\IEEEtitleabstractindextext{%
\begin{abstract}
Despite growing interest in probabilistic modeling approaches and availability of learning tools, people with no or less statistical background feel hesitant to use them. There is need for tools for communicating probabilistic models to less experienced users more intuitively to help them build, validate, use effectively or trust probabilistic models. Users’ comprehension of probabilistic models is vital in these cases and interactive visualizations could enhance it. Although there are various studies evaluating interactivity in Bayesian reasoning and available tools for visualizing the sample-based distributions, we focus specifically on evaluating the effect of interaction on users' comprehension of probabilistic models’ structure. We conducted a user study based on our Interactive Pair Plot for visualizing models’ distribution and conditioning the sample space graphically. Our results suggest that improvements in the understanding of the interaction group are most pronounced for more exotic structures, such as hierarchical models or unfamiliar parameterizations in comparison to the static group. As the detail of the inferred information increases, interaction does not lead to considerably longer response times. Finally, interaction improves users’ confidence.
\end{abstract}

}

\maketitle

\IEEEdisplaynontitleabstractindextext

% \fi
\IEEEpeerreviewmaketitle

\IEEEraisesectionheading{\section{Introduction}\label{sec:introduction}}

\begin{figure*}[!t]%
    \centering
    \subfloat[]{%
        \includegraphics[width=0.2\textwidth]{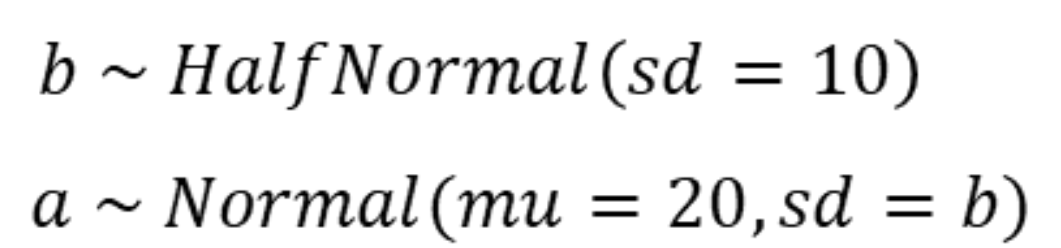}}\quad
    \subfloat[]{%
        \includegraphics[width=0.3\textwidth]{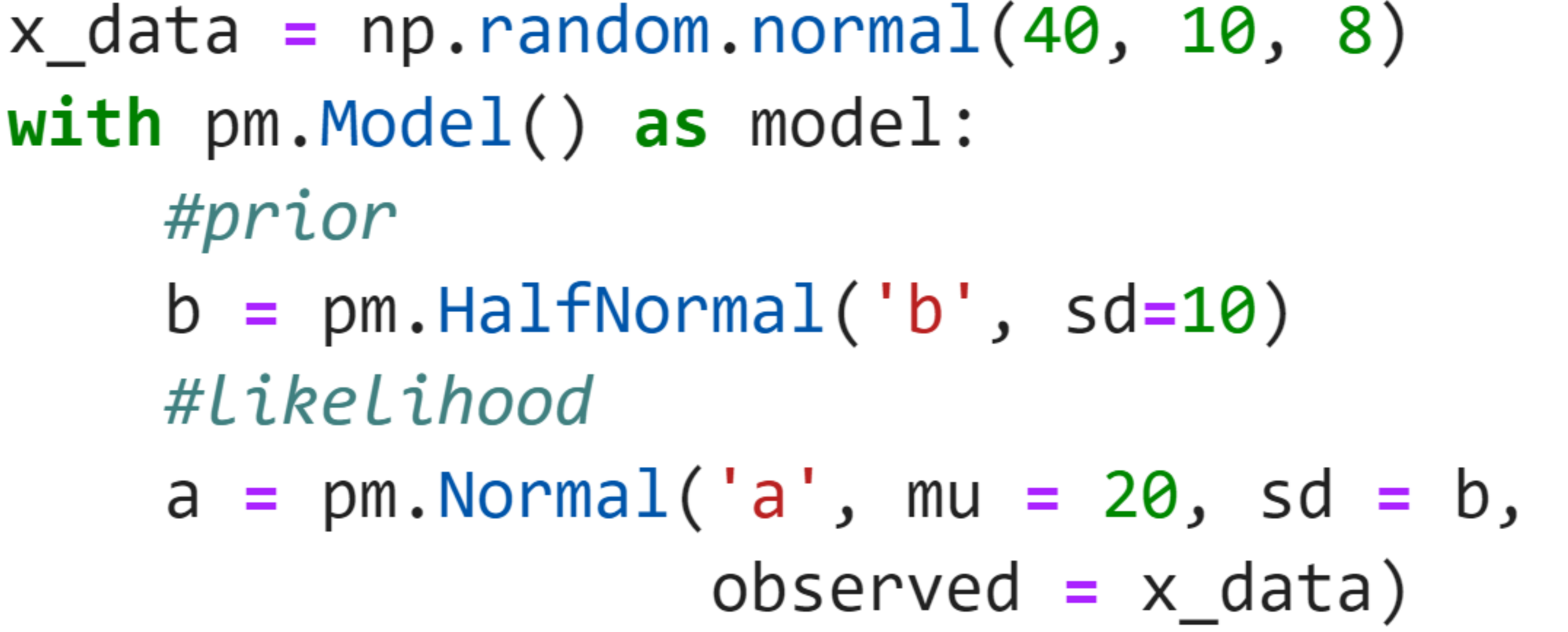}}\quad
    \subfloat[]{%
        \includegraphics[width=0.3\textwidth]{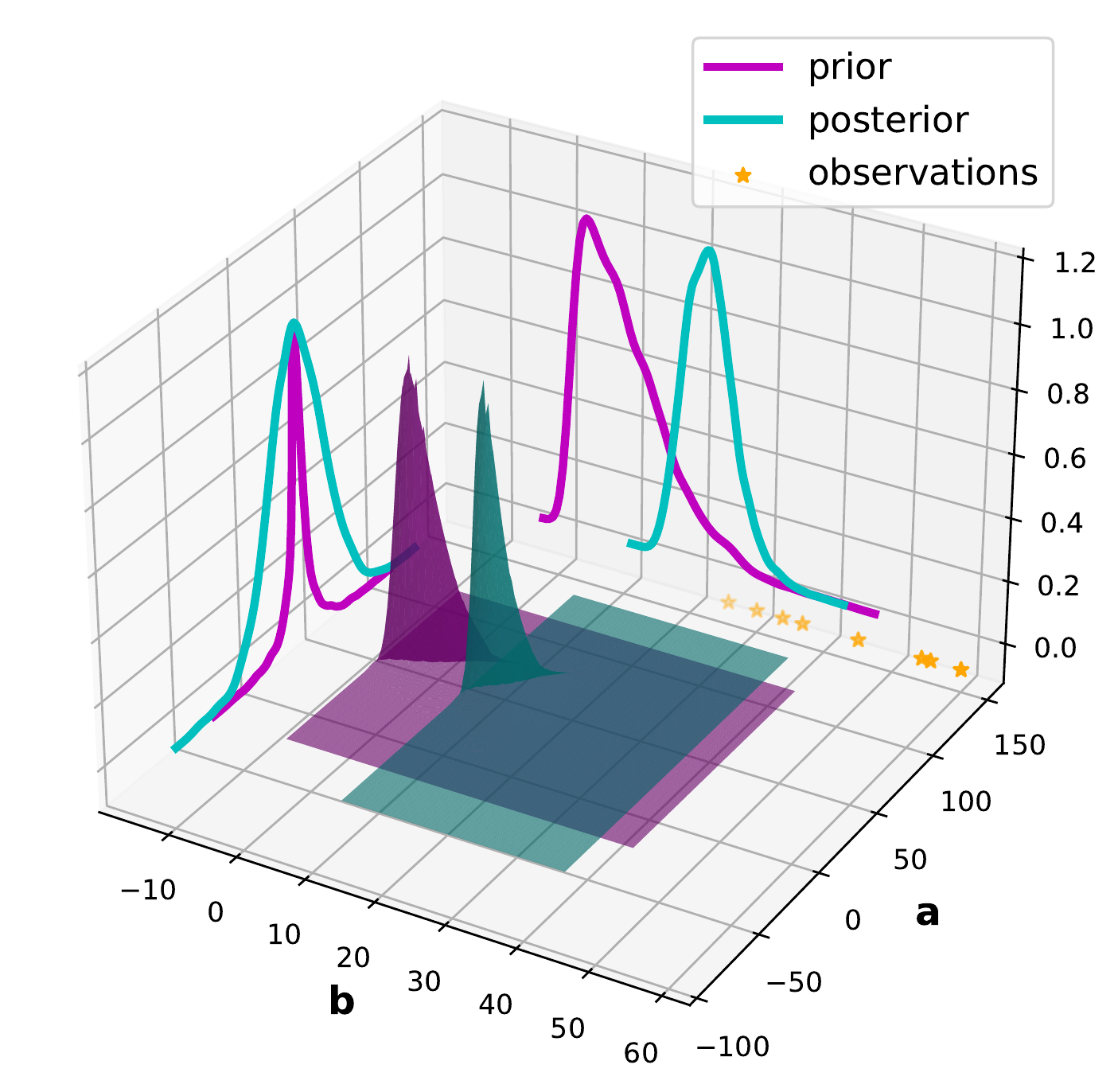}}\\
     \subfloat[]{%
        \includegraphics[width=0.06\textwidth]{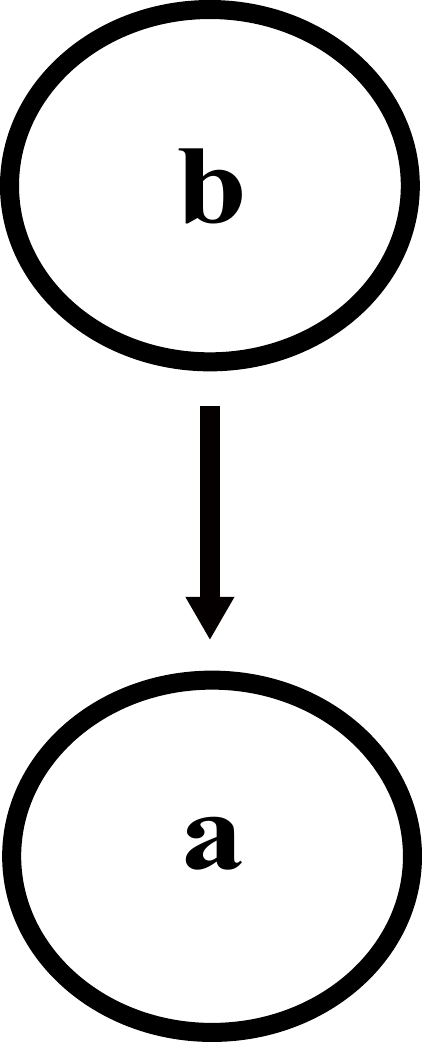}}\quad
    \subfloat[]{%
        \includegraphics[width=0.18\textwidth]{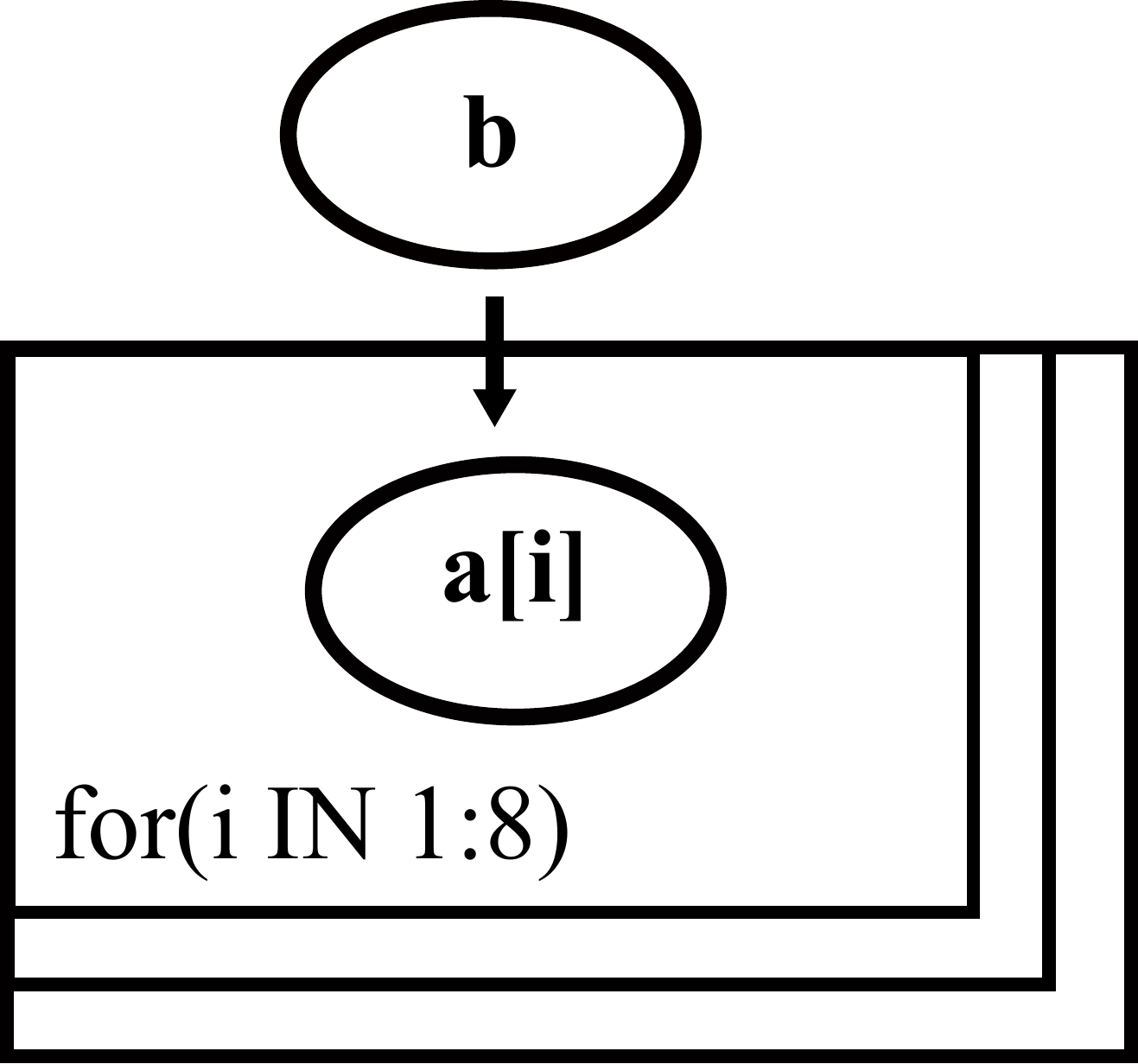}}\quad
    \subfloat[]{%
        \includegraphics[width=0.15\textwidth]{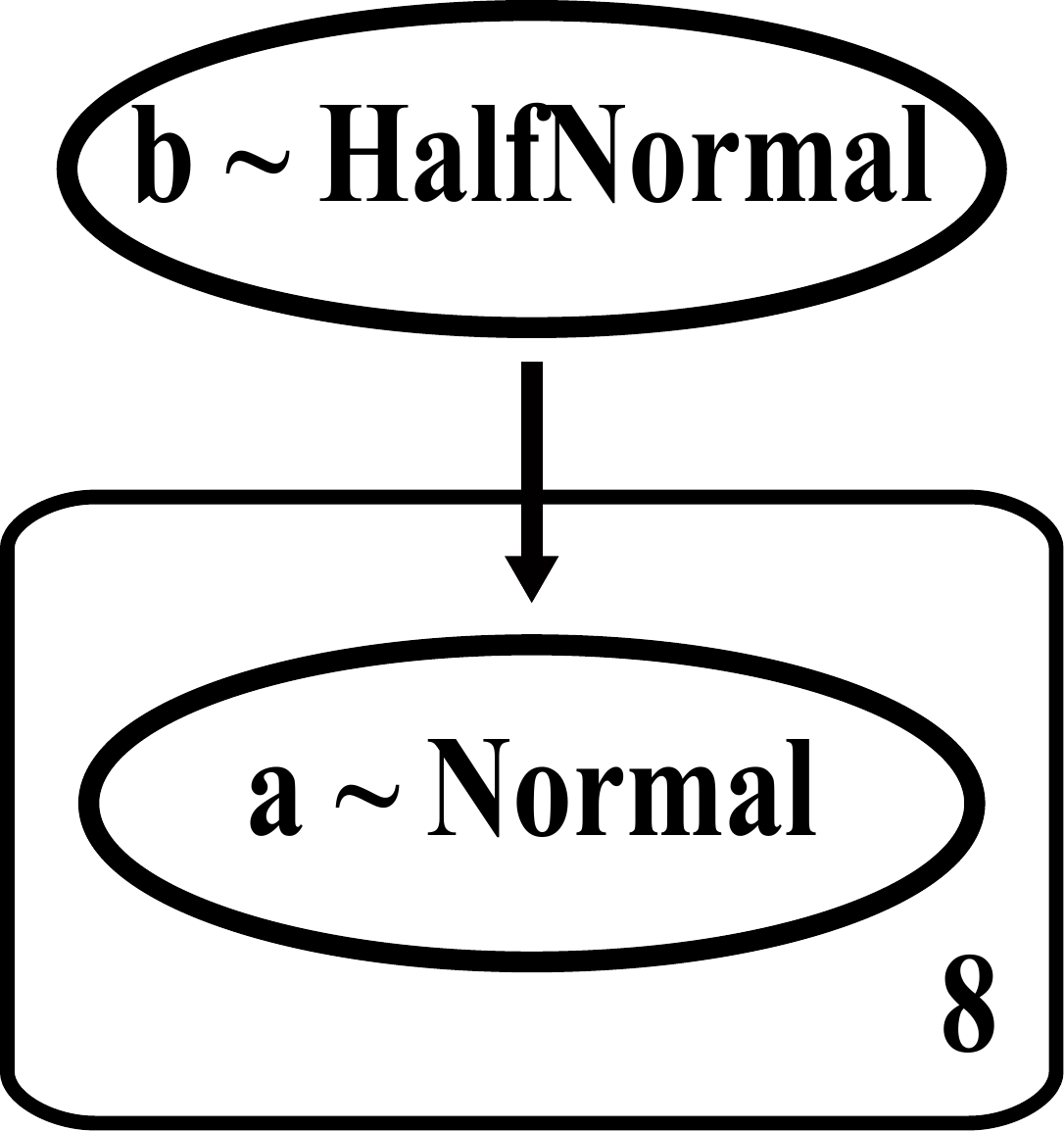}}\quad
    \subfloat[]{%
        \includegraphics[width=0.11\textwidth]{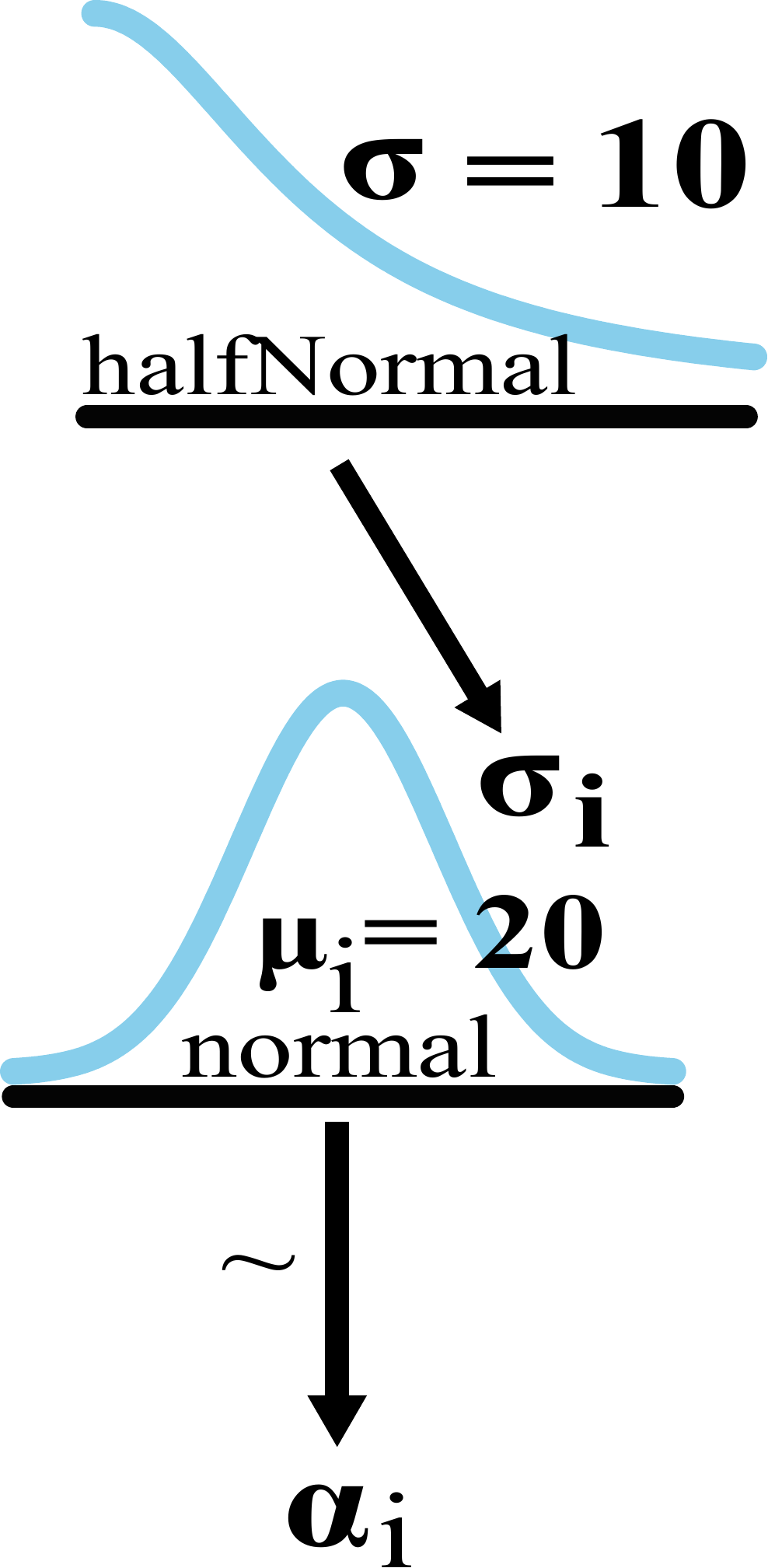}}\quad
    \subfloat[]{%
        \includegraphics[width=0.26\textwidth]{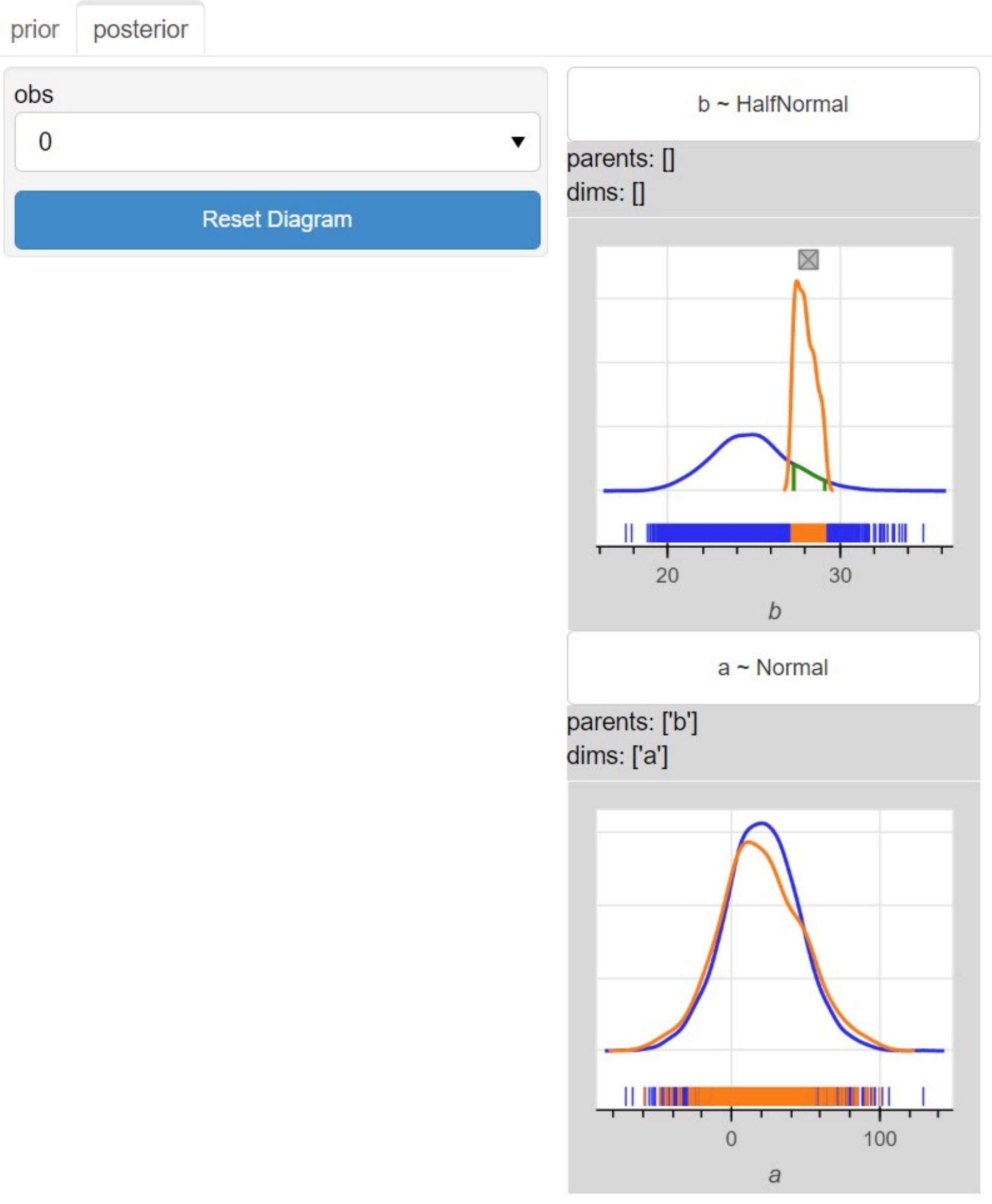}}%
    \caption{Different representations of a probabilistic model. (a) Mathematical definition in probabilistic statements of a simple two-variable model; \(\operatorname{a}\) has a normal likelihood with fixed mean and half-normally distributed standard deviation. (b) The model expressed in a PPL code (PyMC3). (c) The prior and posterior joint (3D surface plot) and marginal distributions (line plots on cube edges) of the model. The posterior is the update of the prior in the light of the observations (yellow stars). (d) Bayesian network, (e) DoodleBUGs' graph, (f) PyMC3's graph, (g) Kruschke-style diagram, and (h) IPME. Varying levels of information is conveyed by each representation.}
    \label{fig:prob_model_1}
\end{figure*}

% Statistical modeling is used for approximating the data generating processes and allowing inferences about models' variables given the observed data.
\IEEEPARstart{P}{robabilistic} modeling is a form of statistical modeling that has increased in popularity lately, especially since the emergence of Probabilistic Programming Languages (PPLs) (e.g. JAGS, BUGs, Stan, PyMC3). PPLs provide an interface for the definition of probabilistic models, implement efficient and well-tested Markov Chain Monte Carlo (MCMC) sampling algorithms for the inference, and automate the inference through literally the push of a button by hiding the details of the implementation. This made probabilistic modeling accessible to a broader audience including people with less solid statistical background.

% There is need for tools for communicating probabilistic models to this diverse audience of probabilistic modeling. Communication of models is required to enable model builders of any level comprehend models' limitations for building suitable models and checking their validity, or even professionals, who rely on such models to do their job and might be of any statistical background, to make more informed decisions (e.g. decision-makers in healthcare or stock market). The mathematical definition of probabilistic models can be complex, unintuitive and hard to understand not only for novices, but also for experts with stronger statistical backgrounds.

Despite the growing interest in Bayesian probabilistic approaches, these methods are not widely adopted. A reason for this might be that people with no or less statistical background do not feel confident to use these methods even when they have access to learning and exploration tools like code templates that guide Bayesian analysis \cite{phelan2019}. The mathematical definition of probabilistic models can be complex, unintuitive and hard to understand not only for novices, but also for experts with stronger statistical backgrounds. There is need for tools for communicating probabilistic models to less experienced users more intuitively to help them build, validate, use effectively or trust probabilistic models. 

Probabilistic models consist of \textit{observed random variables} representing the observed data, and \textit{latent random variables} representing \textit{latent parameters}. Models' (random) variables are modelled by standard probability distributions (normal, uniform, exponential etc.). Probabilistic models are defined mathematically by sets of \textit{probabilistic statements} (Fig.~\ref{fig:prob_model_1}a) or programming PPL exressions (see example in PyMC3 in Fig.~\ref{fig:prob_model_1}b). Although probabilistic statements and PPL expressions is the most informative way to communicate probabilistic models, users with limited statistical background or ignorance of the specific PPL might not be able to understand the technical and mathematical details of probabilistic models.

For example, a probabilistic model is defined by statements \ref{eq:1}-\ref{eq:3}. Parameter \(\operatorname{b}\) is statistically associated with the observed variable \(\operatorname{a}\) because it controls the $\lambda$ parameter of \(\operatorname{a}\)'s distribution. This is a scale parameter that converges to the precision as $\nu$ parameter (degrees of freedom of student-t distribution) increases. The two random variables are also mathematically associated through an exponential transformation. A layperson might struggle to answer queries like ``How does \(\operatorname{a}\)'s uncertainty change with increasing values of \(\operatorname{b}\)?'' given only these expressions.

\begin{equation} \label{eq:1}
\nu \sim \operatorname{Exp}(\lambda=0.1)\\
\end{equation}
\begin{equation} \label{eq:2}
\operatorname{b} \sim \operatorname{Normal}(\mu = 100, \sigma= 10)\\
\end{equation}
\begin{equation} \label{eq:3}
\operatorname{a} \sim \operatorname{StudentT}\left(\nu=\nu,\mu = 0, \lambda = e^{-2 \operatorname{b}}\right)
\end{equation}

This paper focuses on investigating whether interactive visualizations  enhance users' understanding of models' structure, and form stronger mental models without having to dive into mathematical formulations. Interactive visualizations have broadly been used for the exploration of multi-dimensional data \cite{faith2007,sankaran2018,nguyen201} because they are believed to be able to reveal structure in the data more effectively than static visualizations. They have also been used for prior elicitation from users \cite{sarma2020}. There is less investigation though of interactive visualizations for priors' effects on each other within a statistical model. Taka et al. \cite{taka2020} present one such tool, but without empirical evidence of efficacy.

Probabilistic models are characterized by a multi-dimensional \textit{joint distribution} where dimensions correspond to models' variables. In the Bayesian context, there is a \textit{prior} distribution, encoding prior knowledge before seeing observations, which turns into the \textit{posterior} distribution after observation (Fig.~\ref{fig:prob_model_1}c). Taka et al. \cite{taka2020} present an interactive representation of probabilistic models through slicing marginal distributions (Fig.~\ref{fig:prob_model_1}h). Users can condition on marginal distributions to conduct a form of ``sensitivity analysis'' of models' variables. This could reveal \textit{relations} among variables, namely statistical associations or mathematical transformations or equations.

This work's contribution is a user study investigating whether interactive conditioning of probabilistic models could help users identify the existence of relations among variables, the types of relations (e.g. positive or negative correlation) and more detailed structural information (e.g.  statistical associations or mathematical equation among variables). We test accuracy, speed and confidence of identifying these relations. We used an \textit{Interactive Pair Plot (IPP)}, an interactive scatter matrix presenting both the variables' marginal distributions and the pair plots of every pair of joint samples and contours of their pairwise distributions. IPP integrates the interactive conditioning suggested by Taka et al. \cite{taka2020}.

Our Bayesian analysis of the collected data strongly suggests that interactive visualizations like IPP enhance users' comprehension of probabilistic models' structure in cases of more sophisticated model designs that include hierarchical structures or unrelated variables, which are distributed a priori in unfamiliar ways. Response times of the interactive group differ less from the static one as the level of structural detail to be inferred increases. The confidence of the interactive group about their responses was higher than the static group with the effect being stronger in the cases of inferring lower levels of structural detail.

\section{Background: Visualization of Relations}{\label{sec:realted_work}}

\subsection{Representation of Probabilistic Models}
A common way to represent probabilistic models' structure visually is through \textit{graphs}. The \textit{nodes} correspond to models' random variables. The \textit{edges} are directed arrows from one variable to another indicating the direction of their association. The most minimal graph is the Bayesian network \cite{koller2009} (Fig.~\ref{fig:prob_model_1}d). More informed versions of graphs are provided by the graphical tools of some PPLs. For example, in the DoodleBUGs' \footnote{WinBugs' \cite{spiegelhalter2003} model designing environment} graph, nodes contain information about the dimensions of the variables~\footnote{Random variables in a probabilistic model can be multi-dimensional.} (Fig.~\ref{fig:prob_model_1}e). In PyMC3's \footnote{PyMC3 generates automatically the graph of the defined model through its Graphviz interface \cite{ellson2003}.} graphs, nodes also contain the name of the prototype distribution of the variables (Fig.~\ref{fig:prob_model_1}f). The Kruschke-style diagram \cite{kruschke2015_ch8} (Fig.~\ref{fig:prob_model_1}g) elaborates the graph with the iconic ``prototypes'' of the variables' distribution on each node and annotations for the parameters of distributions (e.g. mu, sigma) being set by other parameters in the model.

Static graphs hide the mathematical details of probabilistic models, while preserving some structural information. Users could at a glance view relations among variables or even exact statistical associations or mathematical equations in the case of the more informed versions of the graphs like Kruschke diagrams. But inferring the strength or types of relations (e.g. positive or negative correlations) is still very much dependent on the ability of the users to understand the mathematical model and this becomes harder as variables become more distant in deeply nested hierarchical models. To convey this information visually, we need to communicate conditional distributions of variables. IPME \cite{taka2020} (Fig.~\ref{fig:prob_model_1}h) incorporates the actual samples' distribution into the display of the graph nodes and allows interactive conditioning of the variables to feature relations among them. 

Graphical representations of probabilistic models might be more eloquent in presenting the structure of models in comparison to probabilistic statements and PPL model definitions. But graphs with many variables, levels of hierarchy, or statistical and mathematical details included could become difficult for users to understand. This work investigates whether interactive visualization of probabilistic models' sample-based distribution could help users infer structure more intuitively.

\subsection{Visualization of Inference}
There are existing tools for visualizing probabilistic models' sample-based inference statically or interactively; ArviZ \cite{kumar2019} and IPME \cite{taka2020} in Python, and bayesplot \cite{bayesplot}, tidybayes \cite{tidybayes}, shinystan \cite{shinystan} in R (see review of them in \cite{taka2020}). The following two subsections explain how existing visualizations of sample-based distributions convey relations among probabilistic models' variables.

\begin{figure*}[!t]%
    \centering
    \subfloat[]{%
        \includegraphics[width=0.2\textwidth]{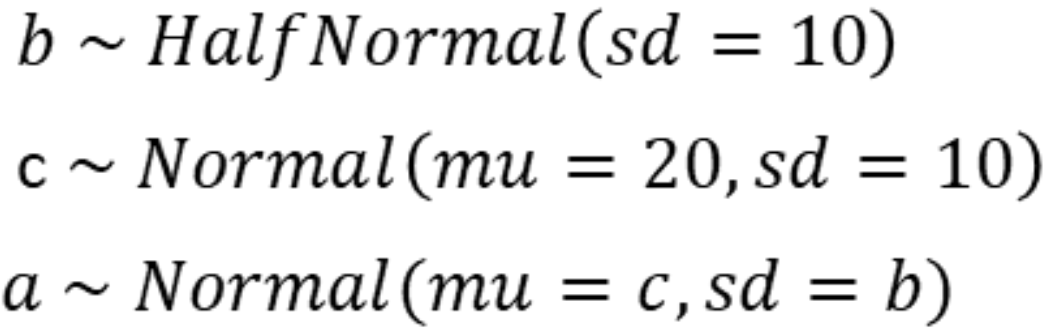}}\\
    \makebox[0.45\textwidth][c]{\parbox[t]{0.45\textwidth}{\textbf{How does \(\operatorname{b}\)'s uncertainty change with \\ increasing values of \(\operatorname{c}\)?}\\ \\ Continuous Dense Conditioning}}\hfil
    \makebox[0.45\textwidth][c]{\parbox[t]{0.45\textwidth}{\textbf{How does \(\operatorname{a}\)'s uncertainty change with \\ increasing values of \(\operatorname{c}\)?}}}\\
    \subfloat[]{%
        \includegraphics[width=0.45\textwidth]{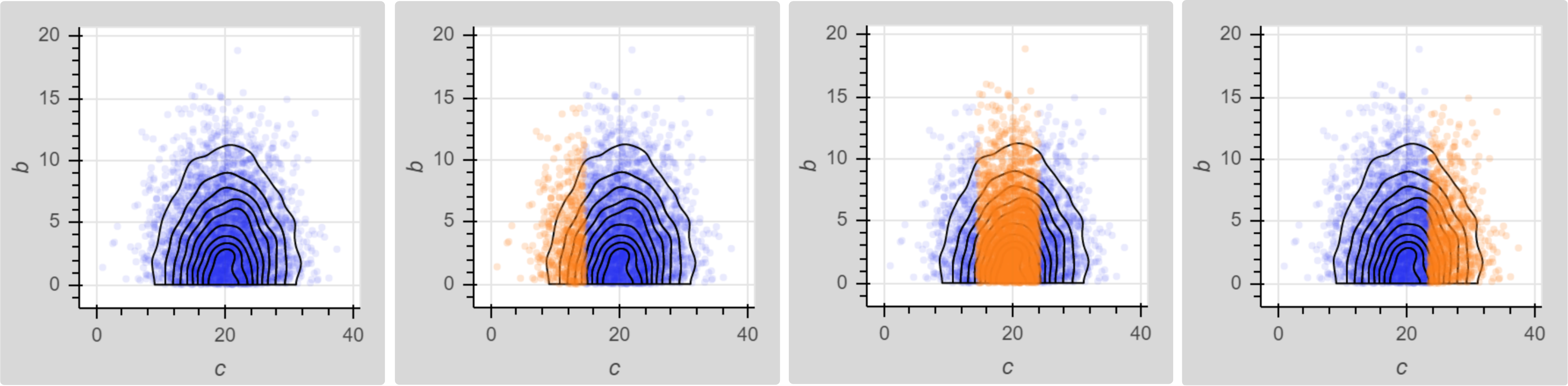}}\quad
    \subfloat[]{%
        \includegraphics[width=0.45\textwidth]{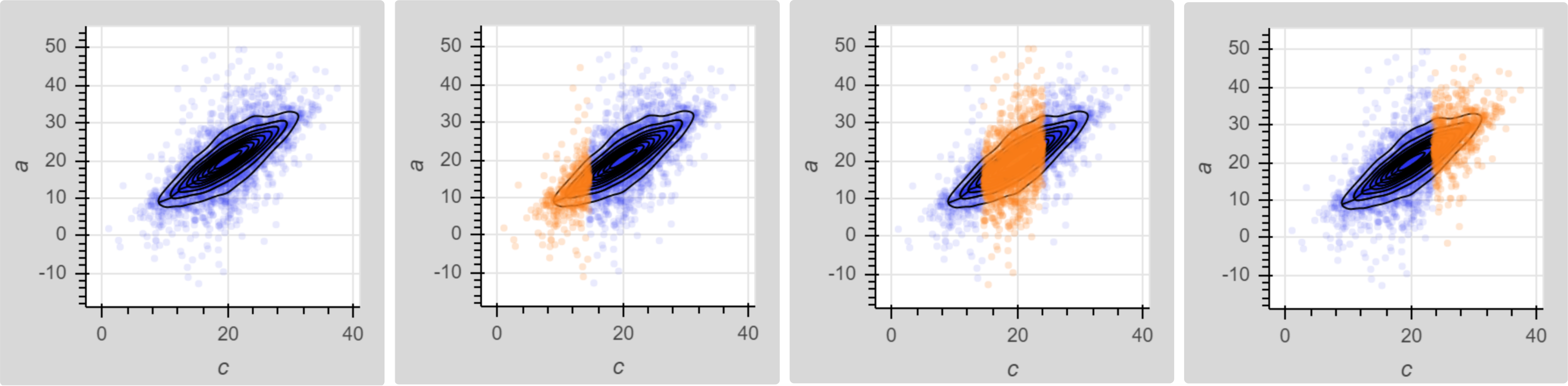}}\\
     \subfloat[]{%
        \includegraphics[width=0.45\textwidth]{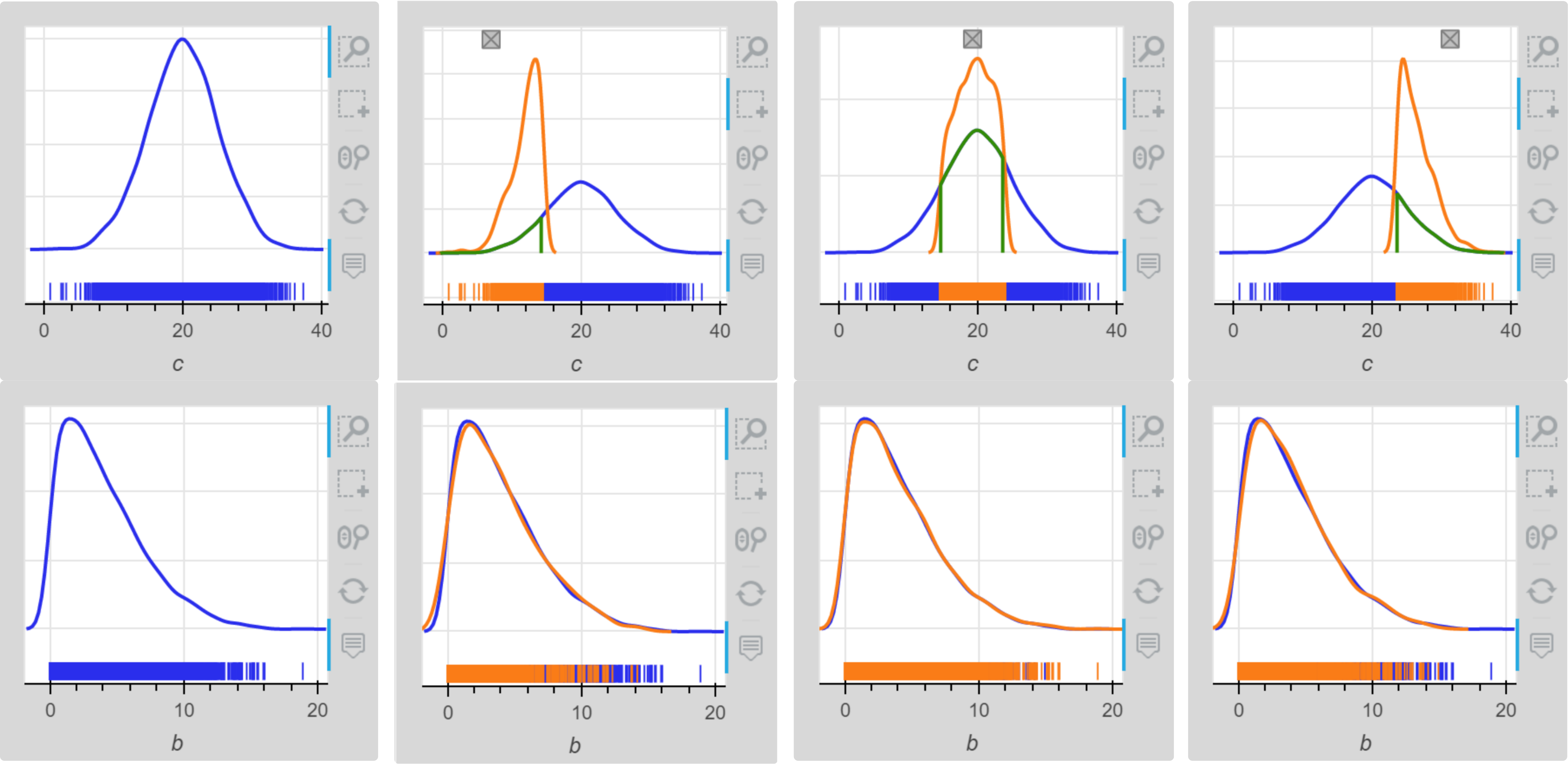}}\quad
    \subfloat[]{%
        \includegraphics[width=0.45\textwidth]{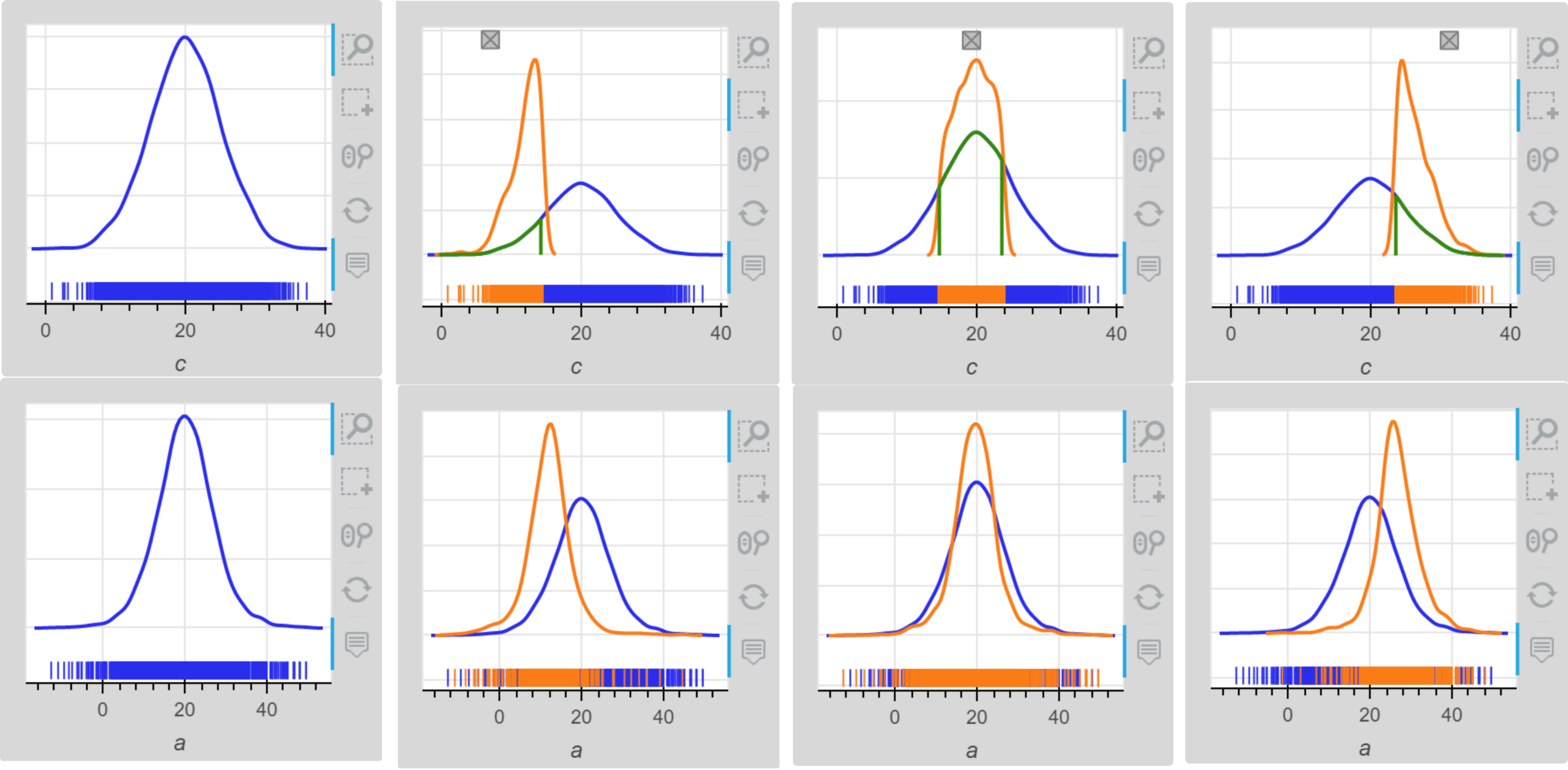}}\\
    \subfloat[\empty]{%
    \makebox[0.93\textwidth][c]{\parbox[t]{0.93\textwidth}{Discontinued Sparser Conditioning}}}\\
    \subfloat[]{%
        \includegraphics[width=0.45\textwidth]{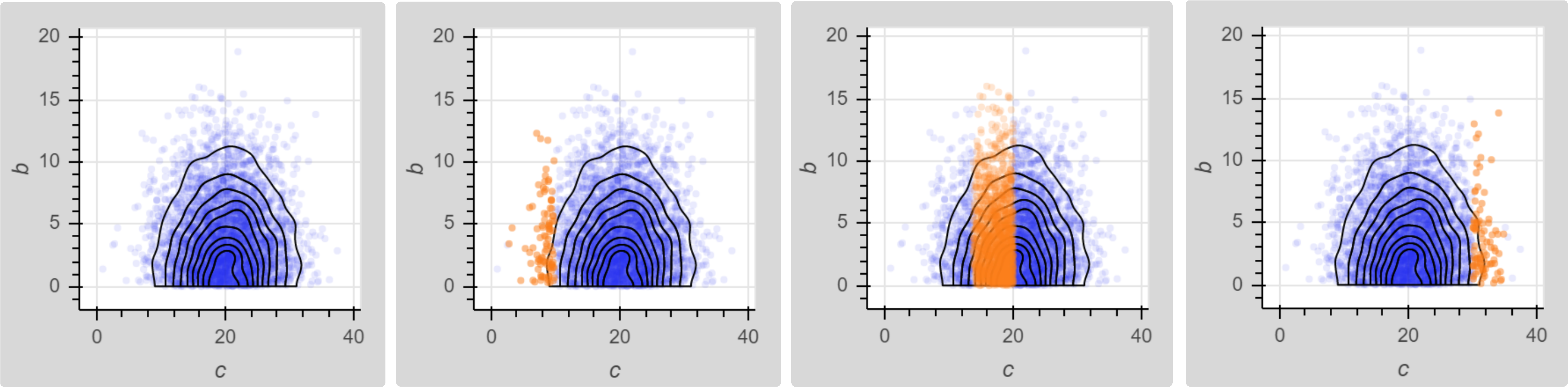}}\quad
    \subfloat[]{%
        \includegraphics[width=0.45\textwidth]{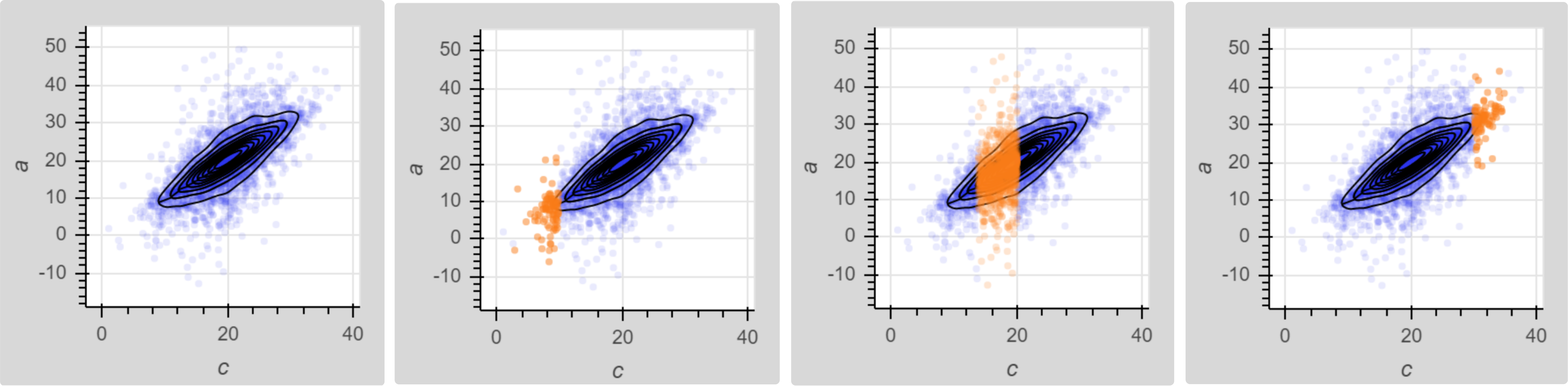}}\\
    \subfloat[]{%
        \includegraphics[width=0.45\textwidth]{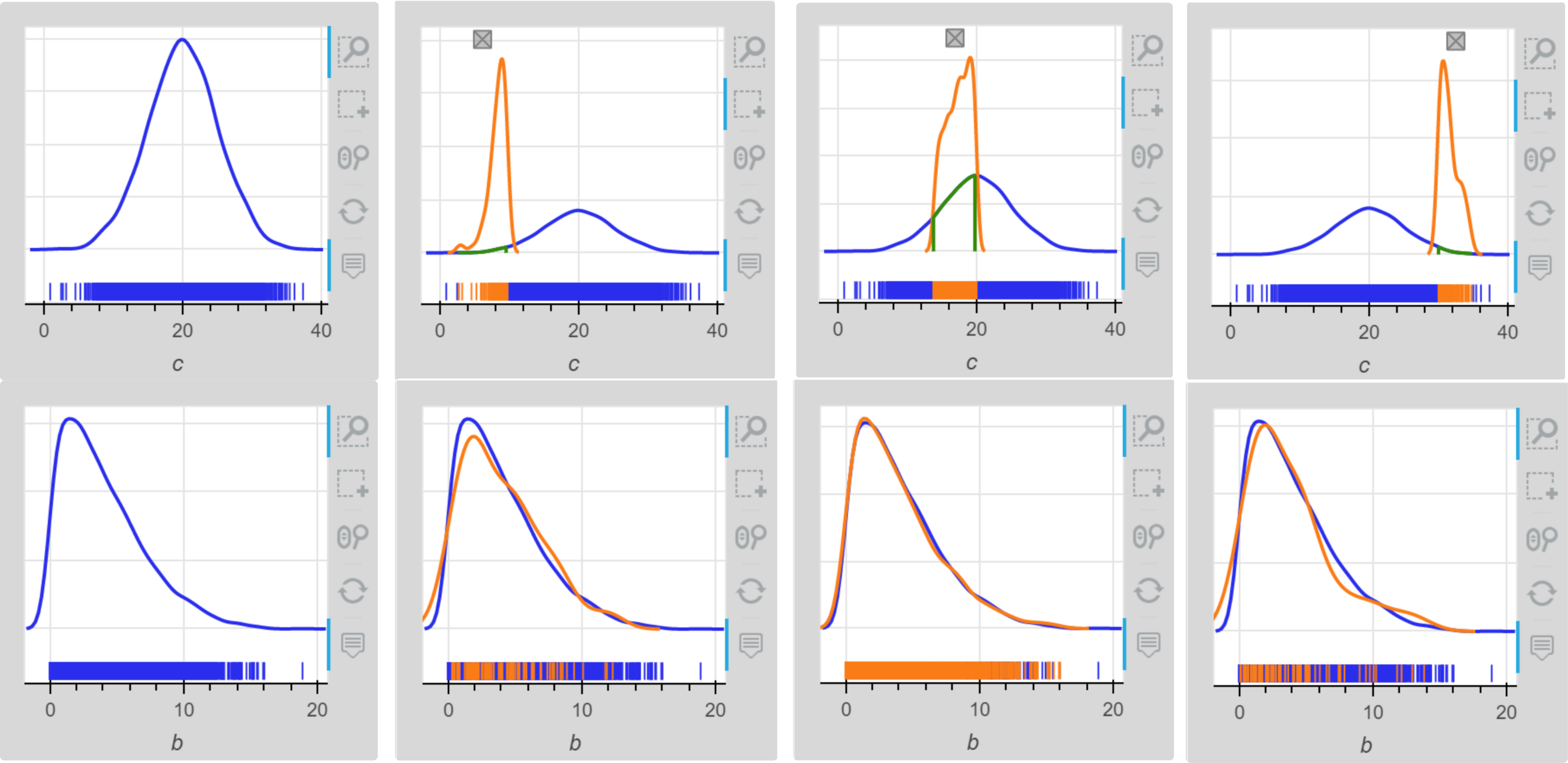}}\quad
    \subfloat[]{%
        \includegraphics[width=0.45\textwidth]{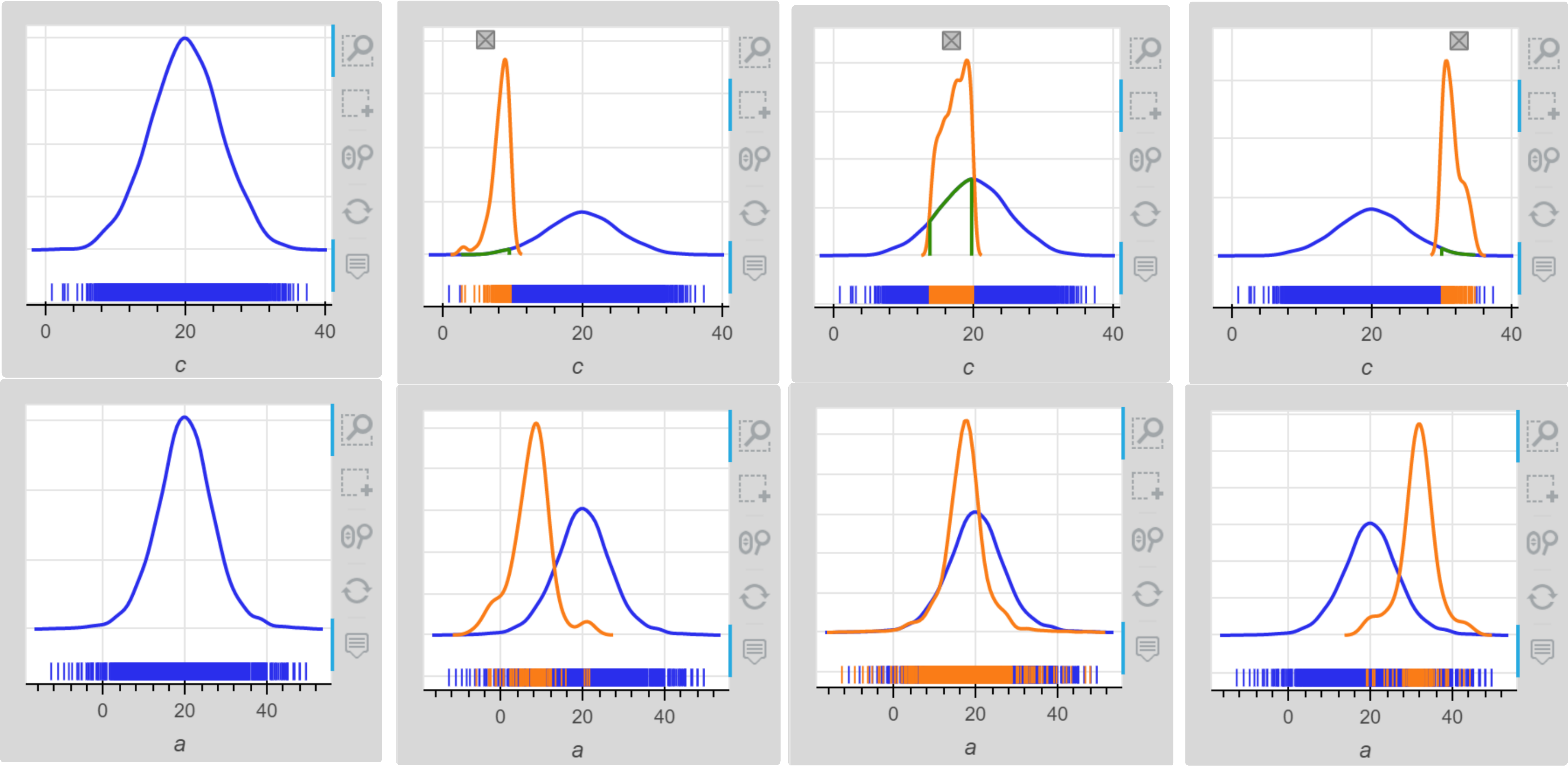}}%
    \caption{Inference-based visualizations of relations among variables of the probabilistic model in (a). Pair plots of model's variables are presented in (b), (c), (f) and (g) and their corresponding marginal distributions in (d), (e), (h) and (i) with instances of sequential conditioning. The figures on the right correspond to unrelated variables, while the figures on the left to related variables. Two different conditioning strategies are presented; (b)-(e) with continuous dense ranges and (f)-(i) with discontinued and varying-density ranges.}
    \label{fig:prob_model_2}
\end{figure*}

\subsubsection{Static Visualization of Relations}
ArviZ Point Estimate Pairplot (APEP) \footnote{\url{https://arviz-devs.github.io/arviz/examples/plot_pair_point_estimate.html}.} presents variables' joint samples and contours of the pairwise distributions on a scatter matrix. This view could enable the inference of relations (correlations) among variables at a glance based on the shape of the pair plots. For example, the well-elongated elliptical shape of the pair plot of \(\operatorname{a}\) and \(\operatorname{c}\) variables in Fig.~\ref{fig:prob_model_2}c implies the existence of a relation. The shape of the pair plot depends on the strength of correlations, the configuration of the 2D Kernel Density Estimation (KDE) algorithm, and the KDE approximation and sampling error. These factors might make the interpretation of pair plots' shape tricky. For example, the shape of the pair plot of variables \(\operatorname{b}\) and \(\operatorname{c}\), which are unrelated, appears conic in Fig.~\ref{fig:prob_model_2}b. This shape might falsely imply the existence of a relation because the dispersion of \(\operatorname{b}\)'s samples seems to decrease at smaller or bigger values of \(\operatorname{c}\); a phenomenon attributed to the finity of the sampling. 

Interpreting pair plots' shape in terms of conditioning could help to resolve these ambiguities. But this could be dependent upon the conditioning strategy applied. For example, conditioning \(\operatorname{c}\) in increasing continuous dense ranges showcases that the variance and mean of \(\operatorname{b}\)'s samples does not change and the mean of \(\operatorname{a}\)'s samples increases in Fig.~\ref{fig:prob_model_2}b. The conditioning strategy (e.g. continuous or discontinued, denser or sparser ranges) might affect the certainty of the inferences about variables' relations. For example, the ranges of \(\operatorname{c}\) at the edges in Fig.~\ref{fig:prob_model_2}f and g might imply a decreased dispersion of \(\operatorname{b}\)'s and \(\operatorname{a}\)'s samples, respectively, due to the finity of sampling. 

\subsubsection{Interactive Visualization of Relations}
IPME \cite{taka2020} presents only the marginal distributions of the variables. Static marginal distributions of variables cannot convey any information regarding the relations among variables. This is enabled in IPME through interactive conditioning by drawing selection boxes to restrict the space of variables (brushing). The marginal distributions of all variables within the restricted space are estimated and drawn (in orange color), and the samples in the restricted sample space are highlighted on the rug plots (linking). Interactively conditioning a variable and observing the distribution of another variable in the restricted sample spaces could reveal relations through the changes of the distribution. 

The change of variables' distributions depends on the type of relations (mathematical or statistical dependencies), and gets affected by the KDE approximation, sampling error and conditioning strategy used. For example, conditioning \(\operatorname{c}\) in increasing continuous dense ranges does not affect the distribution of \(\operatorname{b}\) in Fig.~\ref{fig:prob_model_2}d and leads to an increase of the mean of \(\operatorname{a}\)'s distribution in Fig.~\ref{fig:prob_model_2}e. Conditioning in tiny ranges towards the edges where samples are sparser causes slight changes to the shape of the distribution as the KDE estimation gets affected by the sparsity of the samples. The distribution of \(\operatorname{b}\) deviates from the initial shape when conducting such a conditioning in Fig.~\ref{fig:prob_model_2}h and the width of the distribution of \(\operatorname{a}\) seems to be smaller when conditioning the edges. 

The aim of the user study presented in this paper was to investigate whether adding interactive conditioning of the marginals to a static view of an APEP-like visualization would improve users' judgements about variables' relations in terms of accuracy, confidence and speed.

\subsection{Evaluations of Visualization in Bayesian Reasoning}

To our knowledge, there is no previous work in the existing literature regarding the evaluation of the effect of visualizations in the understanding of probabilistic models' structure. There is though much previous work on the effect of visualization in Bayesian reasoning, where users had to deal with conditioning tasks. Diagrams and contingency tables were found to improve the performance of people in Bayesian reasoning tasks when they were used in the training of the participants in Bayesian reasoning \cite{cole1989}. In another study, frequency representations when used in teaching Bayesian reasoning, had a higher immediate learning effect to learners, and this effect lasted for longer in contrast to training learners in inserting probabilities in Bayes' rule \cite{sedlmeier2001}. 

Brase and Gary \cite{brase2009} conducted a series of experiments and found that people who were using iconic pictorial representations in Bayesian reasoning tasks had signiﬁcantly better performance as compared to people who were using either pictorial representations in the form of continuous ﬁelds or no pictorial representation at all. Micallef et al. \cite{micallef2012} found that there was a reduction in the errors of estimating probabilities based on Euler diagrams, or frequency grids, when these were including explanatory texts instead of numerical information. Ottley et al. \cite{ottley2012} expanded the sample of the study to a more diverse population and found that the results of the previous two papers were not replicated. Ottley et al. \cite{ottley2012}, by conducting the experiments through crouwdsourcing instead of a controlled laboratory environment, demonstrated how sensitive to the crowd the results of such studies can be. Ottley et al. \cite{ottley2016} also conducted a series of experiments and showed that text and visualization designs in regards with the amount of information presented to users can have a significant effect on people's accuracy. 

Several studies of interactive visualizations in Bayesian reasoning have also been conducted. Tsai et al. \cite{tsai2011} developed an interactive visualization to help people solve conditional probability problems and showed that ``Bayes-naive'' people  benefited from this visualization. Their performance in Bayesian reasoning was substantially improved. Breslav et al. \cite{breslav2014} investigated why participants perform poorly in answering  conditional probability questions by analyzing their micro-interactions with the interface where the questions were presented. The findings showed the importance  of careful design of micro-interactions in helping users to better perform in such tasks. Khan et al. \cite{khan2018} found that adding interaction to double tree diagrams when these are used to ``capture the double branching structure of a Bayesian problem'', significantly decreased participants' performance in Bayesian reasoning tasks. This could possibly suggest that too much interaction could cause a cognitive overload to users. Mosca et al. \cite{mosca2021} found also that there was no improvement in users' accuracy in Bayesian reasoning tasks when interaction was used.

\section{Evaluation Study}{\label{sec:evaluation_study}}
\subsection{Study's Research Questions}
The leading research question being investigated by this user study is  ``Do interactive visualizations of probabilistic models' sample-based distribution help users better understand the structure of probabilistic models?''. This  overarching question was broken down into three sub-questions, each of which concerned a different level of detail regarding models' structure:

\begin{itemize}
    \item [RQ1] Do interactive visualizations help users identify the existence or not of {\itshape relations} among probabilistic models’ variables
    \item [RQ2] Do interactive visualizations help users identify the {\itshape type of relation} of models’ variables  
    \item [RQ3] Do interactive visualizations help users to infer {\itshape structural information} about models
\end{itemize}
more accurately, faster, and with more confidence?
 
RQ1 investigates the ability of users to identify the existence or absence of relations among models' variables based on the presented visualization. This is the lowest level of detail regarding models' structure. Relations among variables are represented by the edges on models' graphs. Structurally, RQ1 investigates the ability of users to identify the existence or absence of edges on the graphs among nodes corresponding to models' variables.

RQ2 investigates the ability of users to infer more details about the {\itshape types of relations} among variables. In most cases, the {\itshape relations} of models' variables are {\itshape linear}. In such cases, a polarity characterizes the effect of the parameters on the distribution of their related ones; for example, the occurrence of an increase or decrease of the mean (variance) of a parameter’s distribution when the value of a related parameter increases or decreases. This is a middle level of detail regarding models' structure that this study asks participants to infer. 

RQ3 investigates the ability of users to infer the specific structural information regarding the relations that link parameters together based on the presented visualization; for example, the specific statistical association or mathematical equation that links two or more parameters together. This is the highest level of detail regarding models' structure that this study asks participants to infer.

\begin{figure*}[!t]%
  \centering
  \subfloat[]{%
        \includegraphics[width=0.37\textwidth]{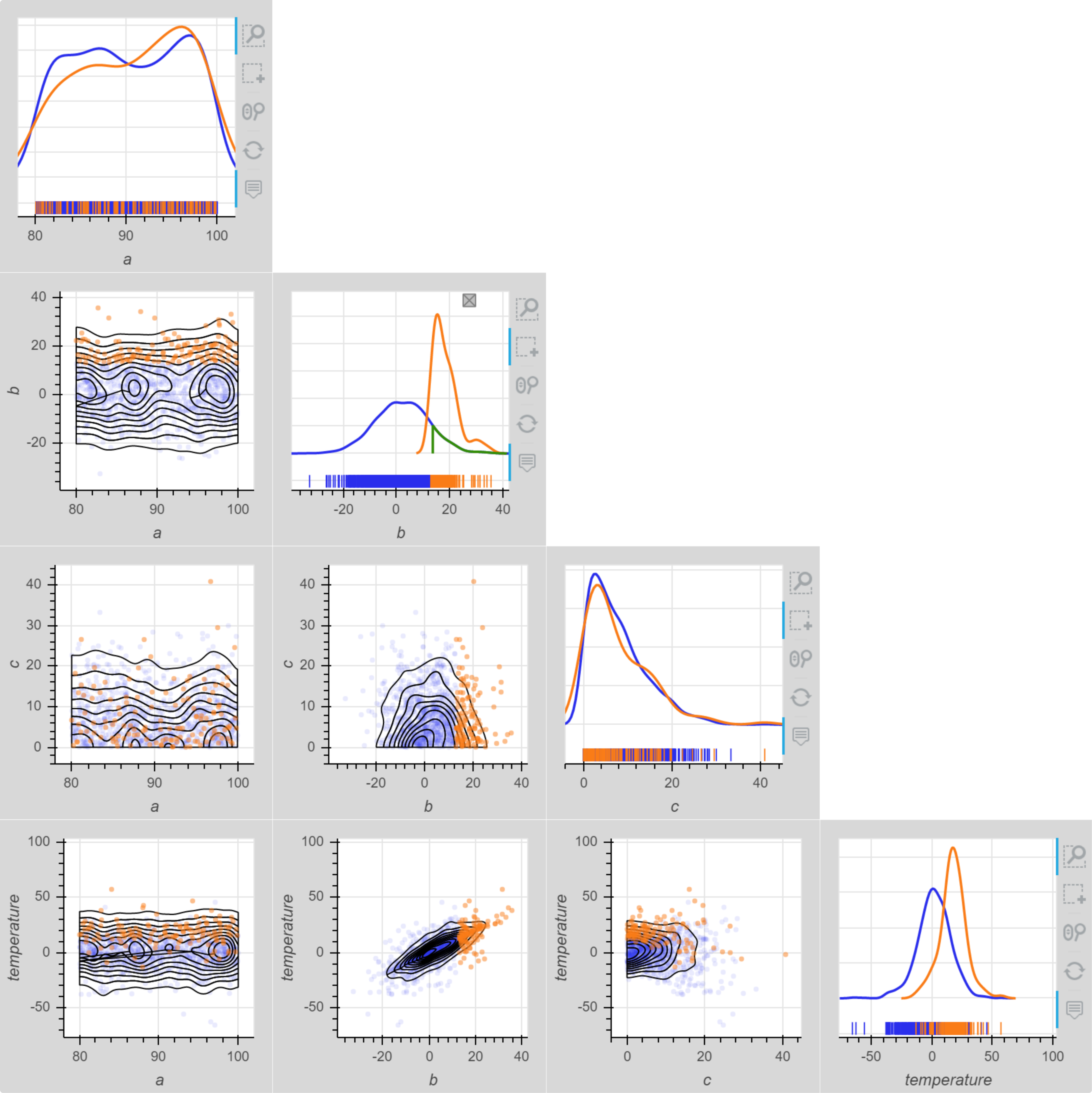}}\quad
  \subfloat[]{%
        \includegraphics[width=0.6\textwidth]{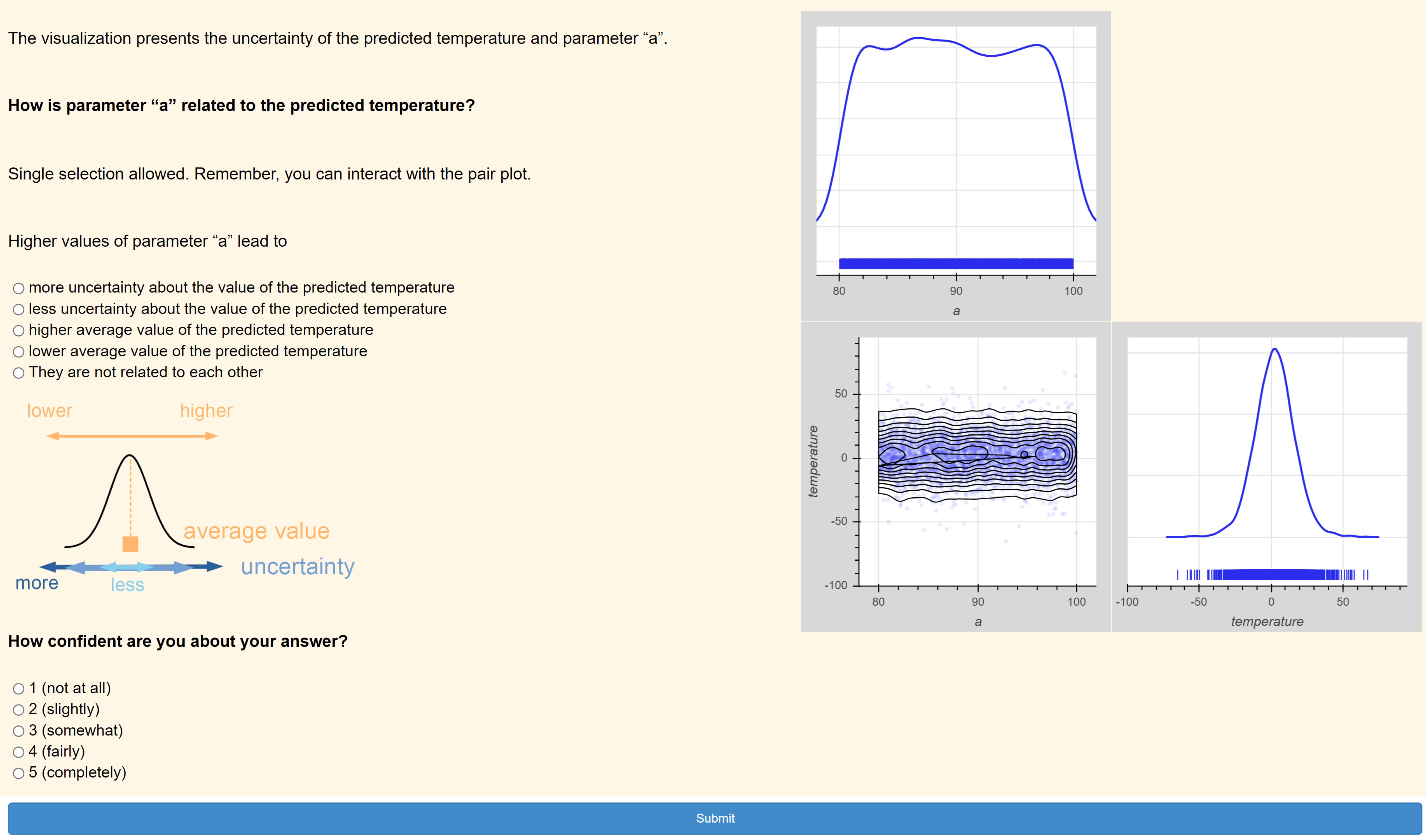}}%
  \caption{(a) Interactive Pair Plot (IPP) of Problem 1 model's inference (see model definition in Table~\ref{tab:questions}) and (b) task t2 (RQ2) as presented to participants.}
  \label{fig:problem1}
\end{figure*}

\subsection{Interactive Pair Plot (IPP)}{\label{sec:interactive_pair_plot}}
\subsubsection{Design of IPP}
The visualization instance used in this user study was IPP\footnote{\url{https://github.com/evdoxiataka/ipme}}. IPP is an interactive scatter matrix for the visualization of the sample-based inference of probabilistic models (Fig.~\ref{fig:problem1}). It was implemented on top of the IPME's framework \cite{taka2020}, and constitutes an extension of IPME by the pair plots of the joint samples of models' variables. The plot cells on the diagonal correspond to models' variables. They present the variables' marginal distributions as a density plot and their samples as a rug plot. The rest of the plot cells across columns or rows present the joint samples of the variables and the contours of their joint distribution. 

The purpose of this user study is to investigate whether users who are using interactive conditioning on the scatter matrix can identify relations and types of relation, and infer more structural details about the models more accurately, faster and with greater confidence in comparison to users who only view a static scatter matrix.

For the scope of the user study, probabilistic models' distribution was presented in the prior space. Models' prior distribution reflects directly their structure. As observations come into models and the prior beliefs are updated, the initial structure of the models can be overwhelmed in the posterior distribution. For a clearer experimental protocol, we focused on the effect of interactive conditioning in the \textit{prior} space on users' understanding of models. The investigation of the effect of observations in the posterior on users' comprehension of models' structure could constitute the subject of a future study. 

All irrelevant interactive elements from IPP's initial design (zoom tools, hovering-over tooltips, tabs, drop-down menus) were removed. Only the selection box tool was kept. IPP was presenting the minimum necessary subset of models' variables to the participants in each study question.  

\subsubsection{Limitations of Implementation}
IPP inherits the limitations of implementation from IPME; for example, rerunning inference to get more samples in sub-ranges of model's sample space with few or no samples and multiple conditions on a single variable cannot be performed online. IPP's API considers subsets of variables of interest to deal with the quadratic scaling in area of the pair plot with the number of variables. This feature could be also added to the graphical interface of the tool in the future.

%%table

\subsection{Study's Design and Participants}
\subsubsection{Participants}
The study had two conditions; the \textbf{static} and \textbf{interactive} version of the IPP. A between-subject design was used, and each participant was randomly assigned to one of the two groups; the interaction (IG) and static (SG) group. Twenty-six people participated in the study with half of them in each group. The study was approved in advance by the institution's ethics review board (approval number 300200319). Participants were recruited through mailing lists and social media of the institution without any requirement regarding their statistical background, and were offered a £10 worth online shopping voucher as a compensation for their time. The study was conducted online.

\subsubsection{Study's Structure}
There were three distinct parts in the study; training, study questions, and demographic questions. The training included four videos followed by short discussion to answer participants' questions. The training videos presented the aim and structure of study, an introduction to basic probabilistic concepts (e.g. random variable, probability, density plot, sampling from distribution), an explanation of the assigned version of the IPP, and some example questions similar to the study questions. More details about the training videos can be found in the supplemental material.

The study questions were divided into three parts corresponding to probabilistic models of increasing complexity. A set of questions of all three levels of structural detail (RQs) was created for each model. Table \ref{tab:questions} presents a summary of the models and questions. There were nineteen questions altogether. All participants, independently of condition, answered exactly the same questions. The problems and questions were presented in increasing difficulty and level of structural detail, and in the same order to all participants. The only difference among participants was the version of the IPP.

\begin{figure*}[!t]
    \centering 
    \subfloat[]{%
        \includegraphics[width=0.32\textwidth]{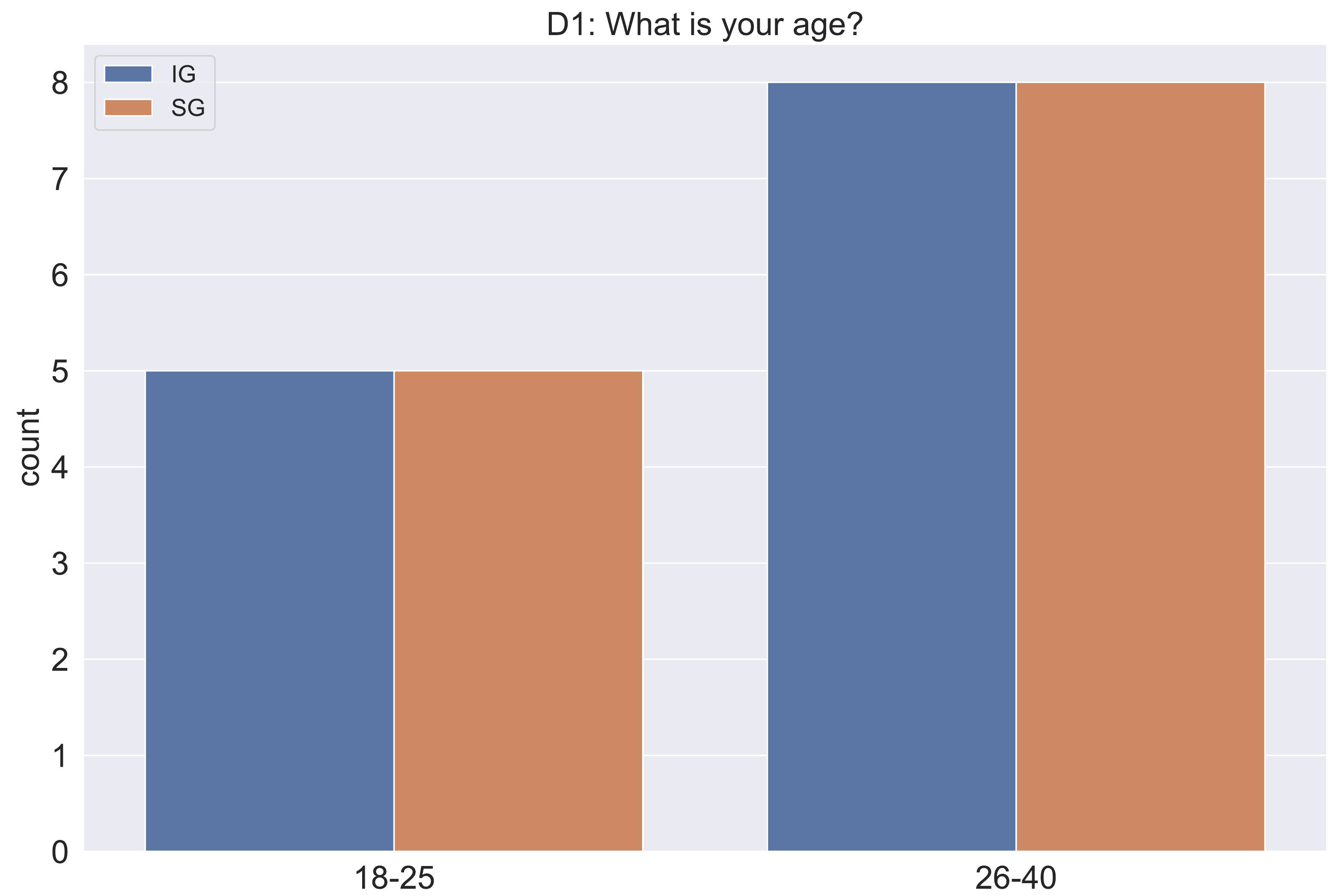}}\quad
    \subfloat[]{%
        \includegraphics[width=0.32\textwidth]{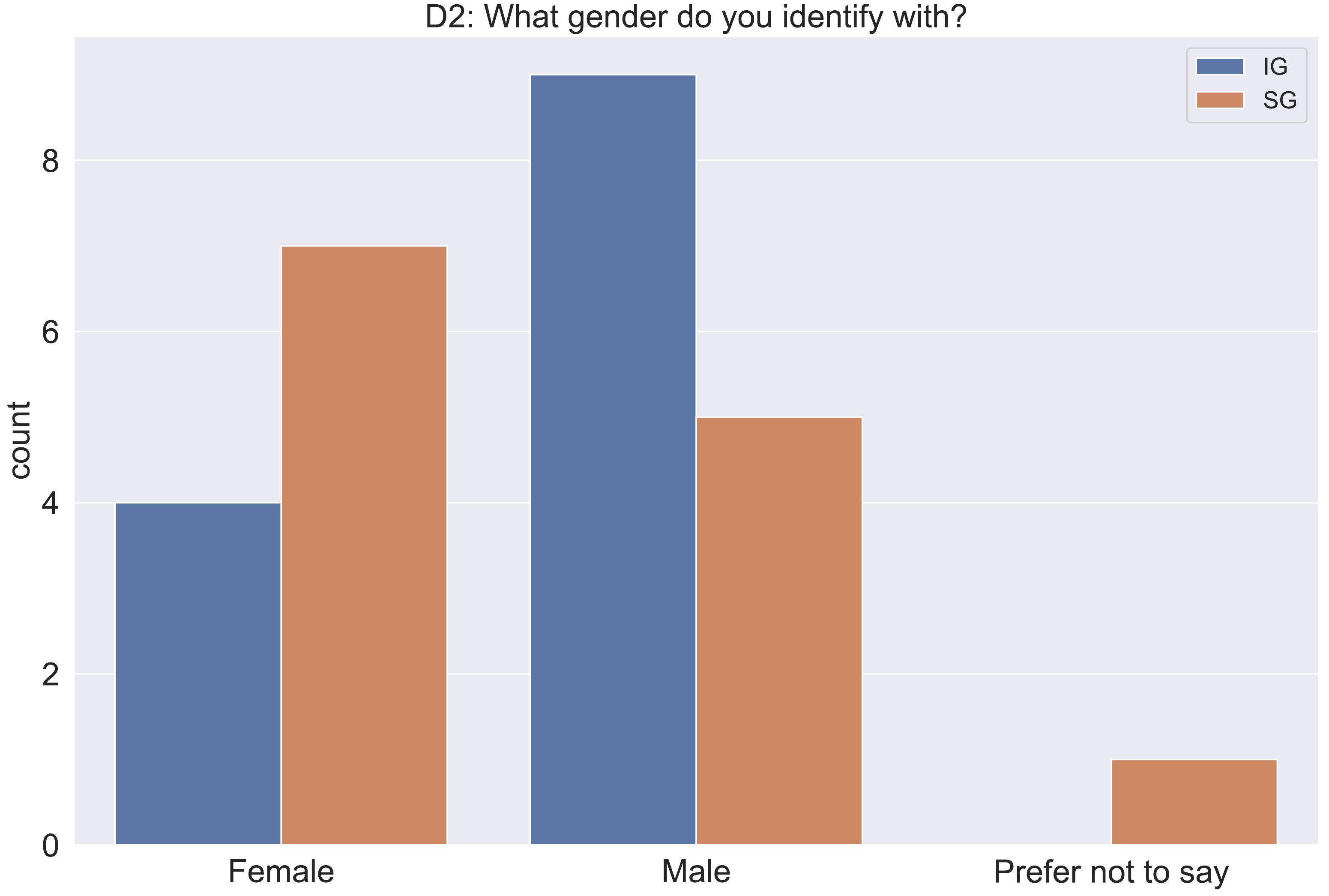}}\quad
    \subfloat[]{%
        \includegraphics[width=0.32\textwidth]{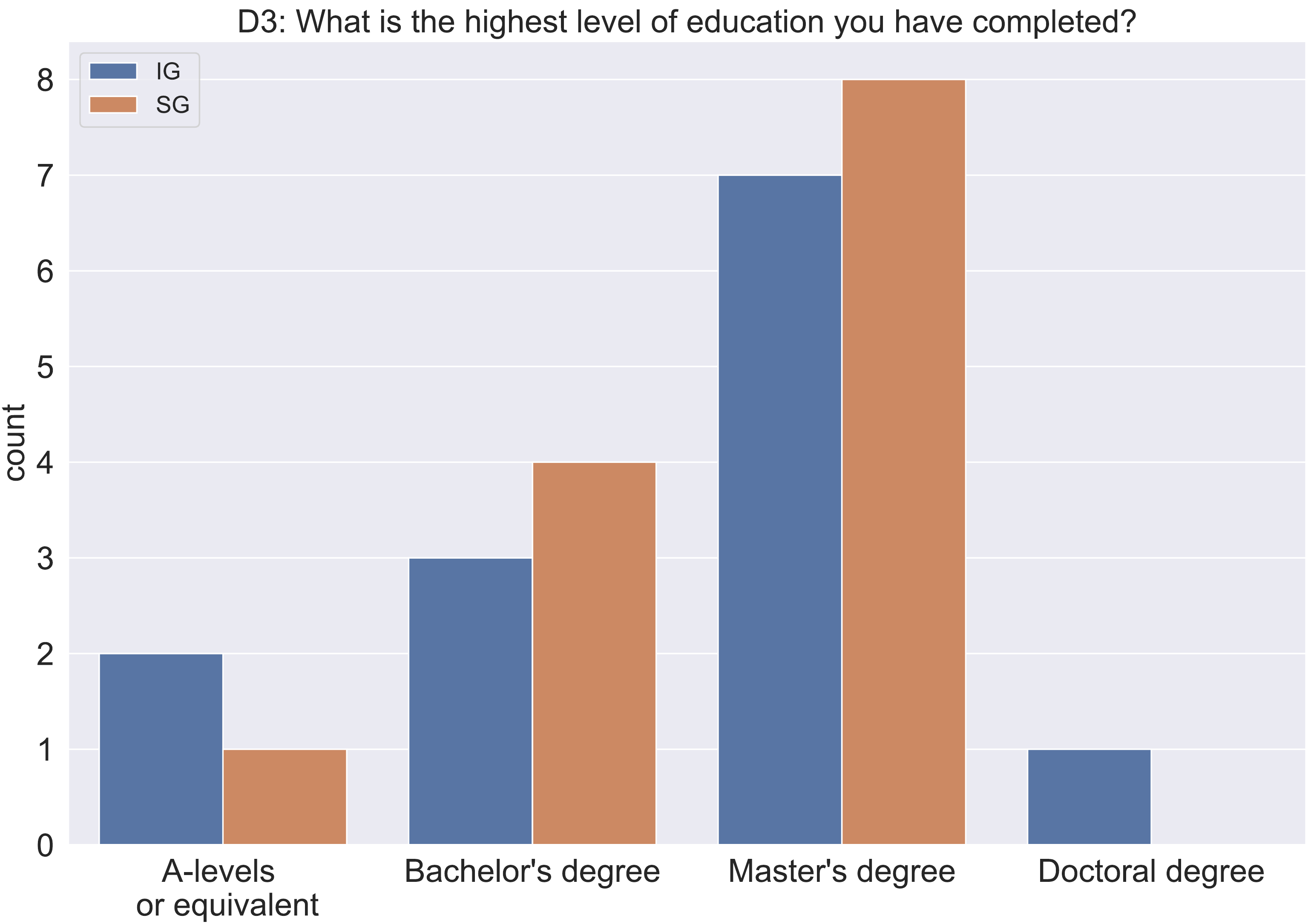}}\\
    \subfloat[]{%
        \includegraphics[width=0.47\textwidth]{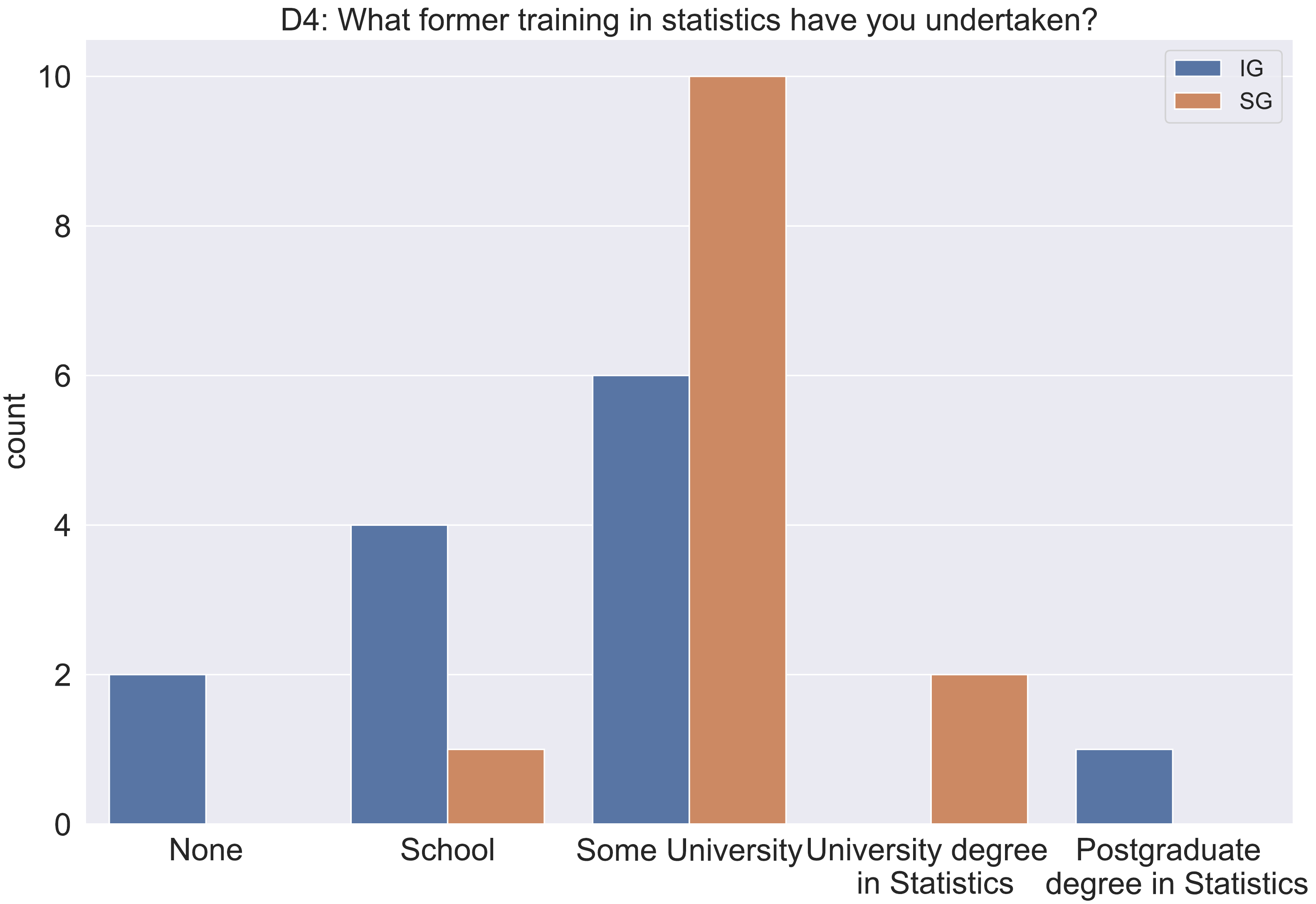}}\quad
    \subfloat[]{%
        \includegraphics[width=0.47\textwidth]{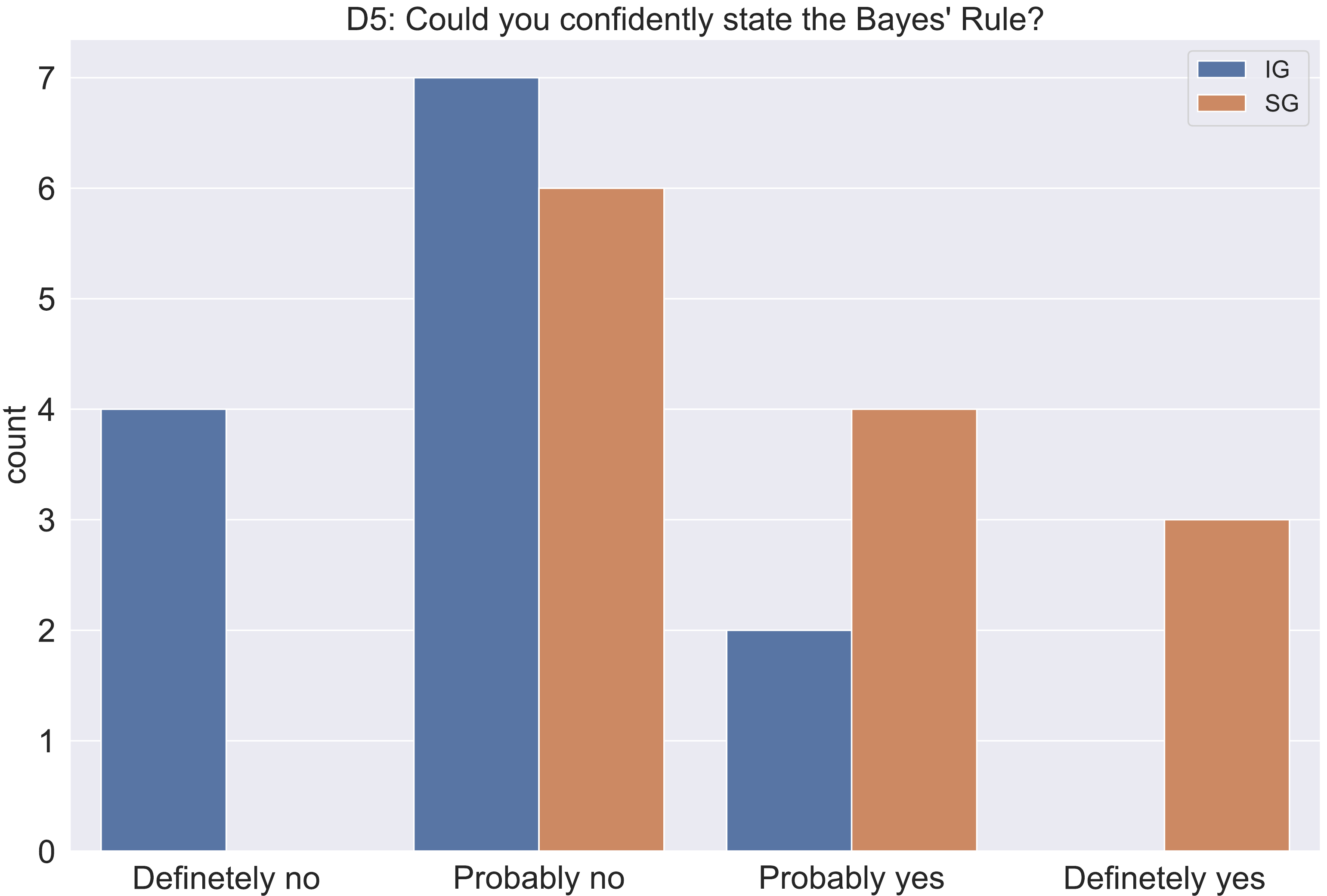}}%
    \caption{Bar graphs presenting the demographic statistics of participants' groups (IG and SG). (a) Age group (b) Gender (c) Highest educational level completed (d) Former training in statistics (e) Confidence to state Bayes' rule bar graphs. Both groups comprised of more older participants. There was a slight gender imbalance between the groups with IG having more males and SG more females. The educational background was generally well-balanced between the groups, while participants in SG had a slightly higher former training in Statistics.}
    \label{fig:demo}
\end{figure*}

At the outset of each trial we captured basic participant demographic information, including the age, gender, highest educational level completed, former training in statistics and knowledge of Bayes' rule. The demographics statistics of the participants is presented in Fig.~\ref{fig:demo}.

\subsubsection{Models' Design}\label{sec:model_design}
Three probabilistic models with increasing complexity were designed for this user study. Each model had an observed random variable with semantically meaningful name (\(\operatorname{temperature}\),  \(\operatorname{random\_number}\), \(\operatorname{reaction\_time}\)) and a set of unidentified parameters named with letters \(\operatorname{a}\),\(\operatorname{b}\),\(\operatorname{c}\) etc. The definitions of the models are presented in Table~\ref{tab:questions}.

\begin{itemize}
    \item \textbf{Problem 1} was the simplest one; a normal likelihood where the unidentified parameters were directly setting the mean and variance of the observed variable. 
    \item \textbf{Problem 2} used a slightly more complex parameterization; a uniform likelihood with the upper and lower bounds set by the unidentified parameters through a deterministic transformation: \(\operatorname{lower\_bound} = a-c\) and \(\operatorname{upper\_bound} = a+c\). 
    \item \textbf{Problem 3} was an hierarchical linear regression model with a normal likelihood, where the mean was set as \(\operatorname{\mu} = a+b*day\) and there were hyper-priors set for the priors of the \(\operatorname{a}\) and \(\operatorname{b}\) parameters.
\end{itemize}

The problems were designed to include a variety of distributions, parameterizations, and strengths of correlation. One of the unidentified parameters in each problem was {\itshape unrelated} to the rest of variables and parameters. We used a variety of prior distributions for the unrelated unidentified parameters; a uniform in Problem 1, a half-normal in Problem 2, and a normal in Problem 3. 

All models were designed and implemented in PyMC3 and the ArviZ library and arviz\_json\footnote{\url{https://github.com/johnhw/arviz_json}} package were used to extract the inference data in the required input format for IPP. The PyMC3 code for the definition of the models can be found in the supplemental material.  

\begin{figure*}[!t]%
    \centering
    \subfloat[]{%
        \includegraphics[width=0.34\textwidth]{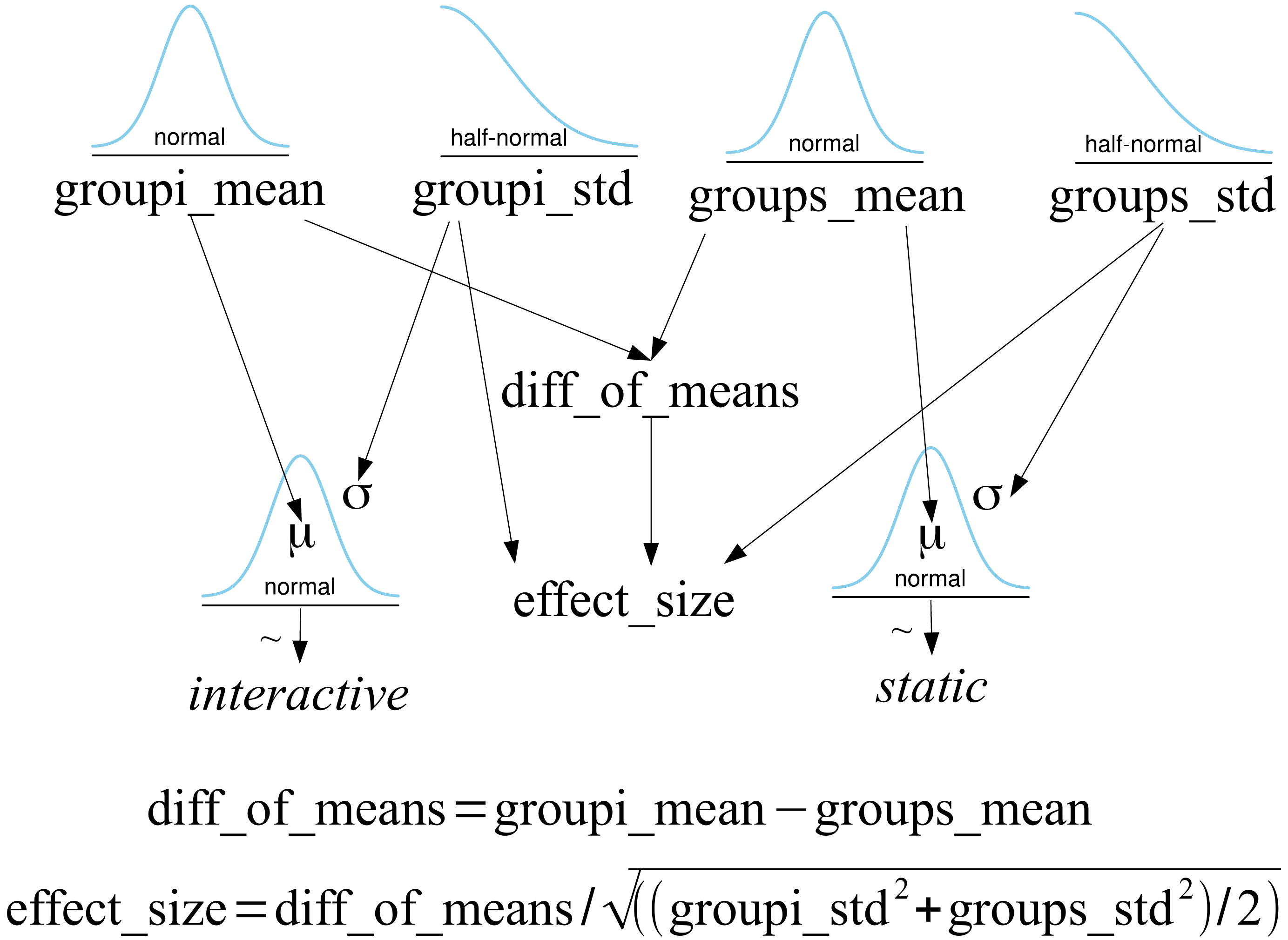}}\quad
    \subfloat[]{%
        \includegraphics[width=0.28\textwidth]{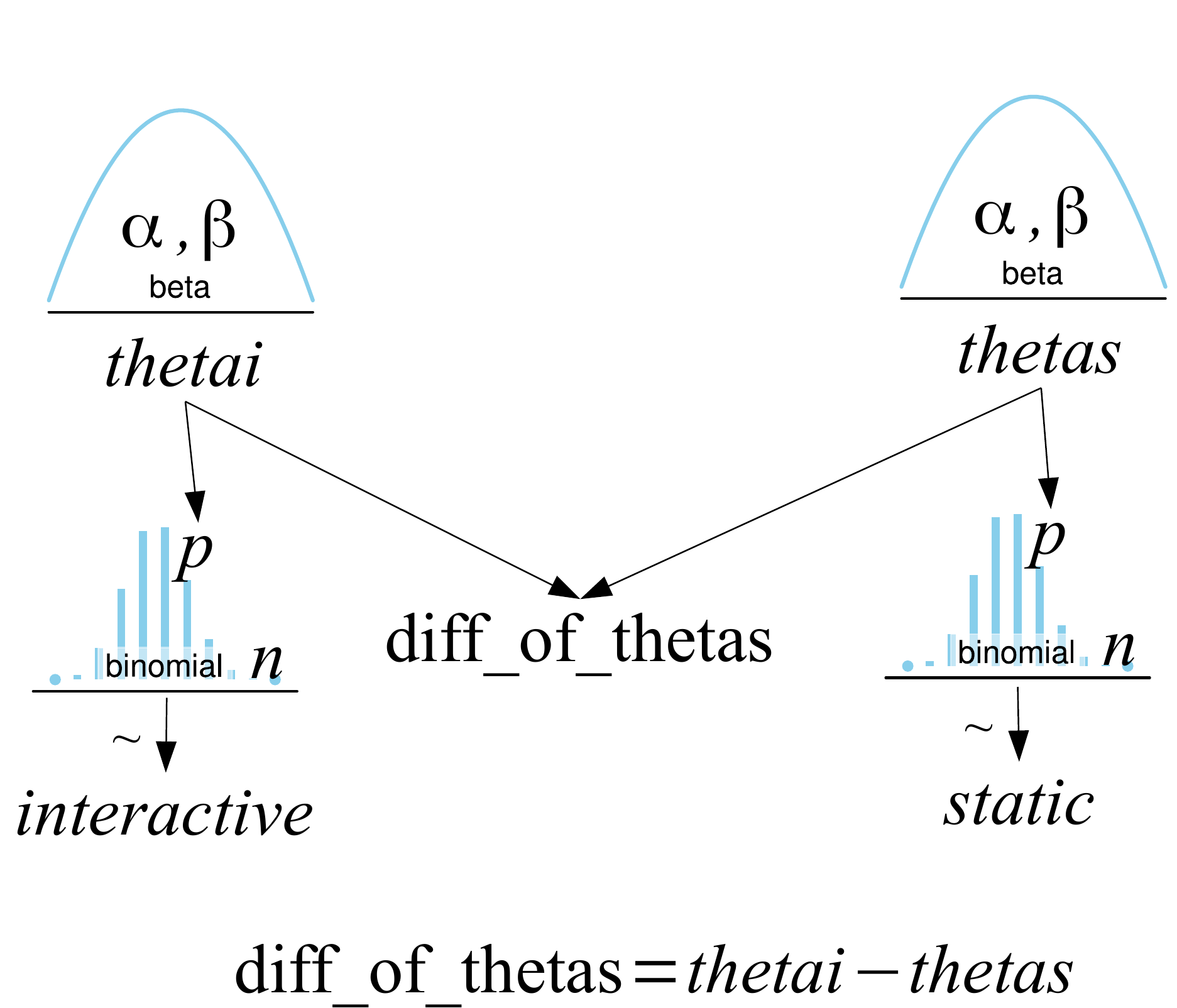}}\quad
    \subfloat[]{%
        \includegraphics[width=0.28\textwidth]{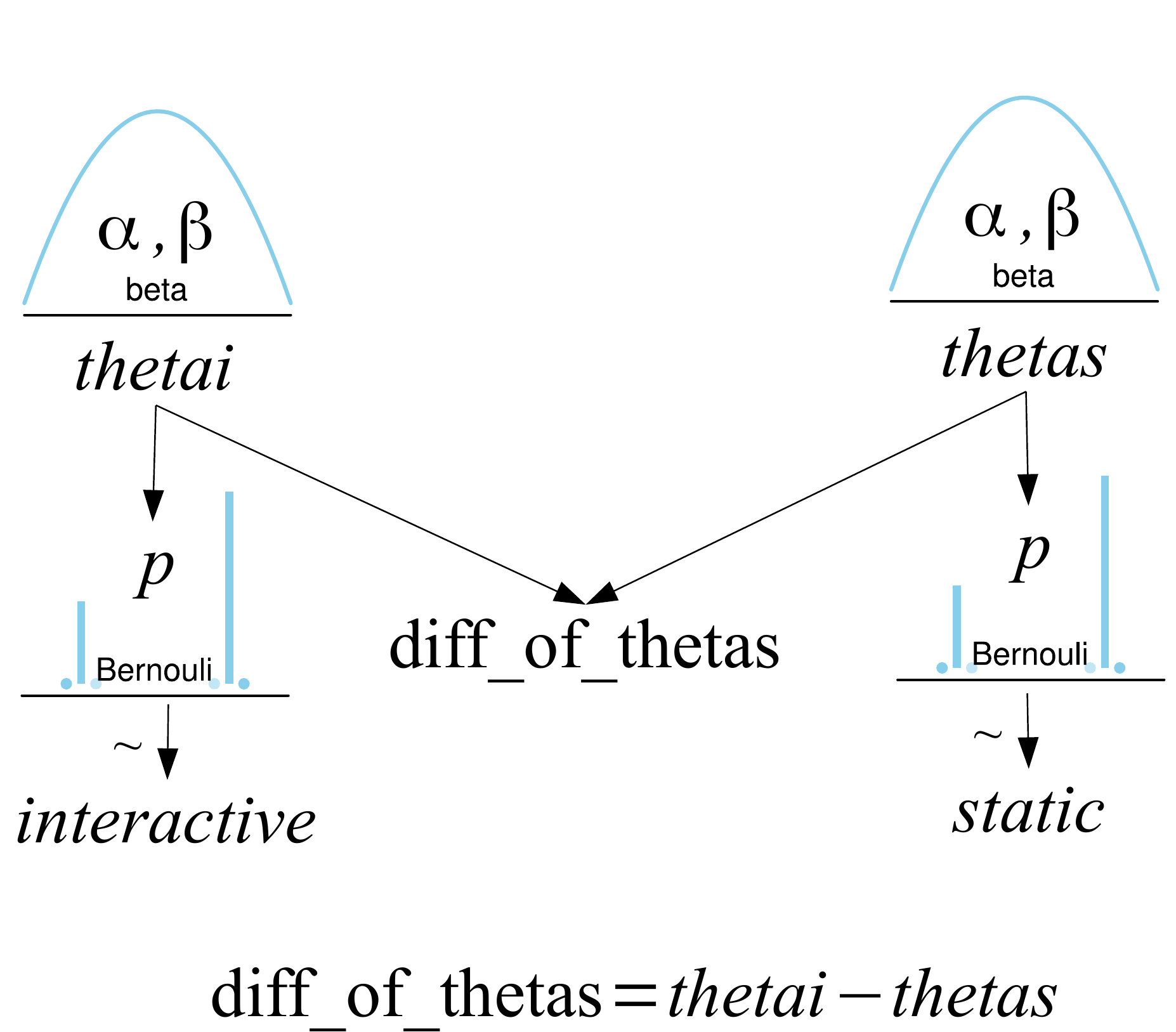}}%
    \caption{Kruschke-style diagrams of the probabilistic models used for the analysis of the (a) response times and confidence, (b) accuracy in RQ1 tasks, and (c) accuracy in RQ2 and RQ3 tasks.}
    \label{fig:analysis_models}
\end{figure*}

\subsubsection{Tasks' Design}\label{sec:task_design}
All questions were multiple-choice. Multiple selections were allowed for the RQ1 questions, and single selection for the rest. Each available option was graphically illustrated in the cases of RQ2 and RQ3 questions. Participants' confidence was input in a five level Likert scale. The following list presents a Problem 1's question for each RQ and Fig.~\ref{fig:problem1} presents the RQ2-t2 question of Problem 1 as presented to participants. A detailed list of the questions can be found in the supplemental material.

 \begin{itemize}
    \item [RQ1.] Which of the parameters ``a'', ``b'' and ``c'', if any, do you think are related to the temperature?
    
    \textbf{Multiple selections allowed.}
    \begin{itemize}
        \item[$\square$] a
        \item[$\square$] b
        \item[$\square$] c
        \item[$\square$] none
    \end{itemize}
    \item [RQ2.] How is parameter ``a'' related to the predicted temperature? 
    
    \textbf{Single selection allowed.}
    
    Higher values of parameter ``a'' lead to 
    \begin{itemize}
        \item[$\square$] more uncertainty about the value of the predicted temperature
        \item[$\square$] less uncertainty about the value of the predicted temperature
        \item[$\square$] higher average value of the predicted temperature
        \item[$\square$] lower average value of the predicted temperature
        \item[$\square$] They are not related to each other
    \end{itemize}
    \item [RQ3.] How would you describe the effect of parameters ``a'', ``b'' and ``c'' on the predicted temperature?
    
    \textbf{Single selection allowed.}
    \begin{itemize}
        \item[$\square$] ``a'' controls the average value, ``b'' the uncertainty and ``c'' has no effect on the predicted temperature
        \item[$\square$] ``a'' controls the average value, ``b'' has no effect and ``c'' controls the uncertainty of the predicted temperature
        \item[$\square$] ``a'' controls the uncertainty, ``b'' the average value and ``c'' has no effect on the predicted temperature
        \item[$\square$]  ``a'' controls the uncertainty, ``b'' has no effect and ``c'' controls the average value of the predicted temperature
        \item[$\square$]  ``a'' has no effect, ``b'' controls the average value and ``c'' the uncertainty of the predicted temperature
        \item[$\square$]  ``a'' has no effect, ``b'' controls the uncertainty and ``c'' the average value of the predicted temperature
        \item[$\square$] There is no effect.
    \end{itemize}
 \end{itemize}

\subsection{Analysis and Results}
\subsubsection{Expected Effects and Measures}

This user study investigated three expected effects by the use of interactive visualizations; accuracy, response time and confidence of the participants. There were three measures that were elicited in this user study to assess whether each of the corresponding expected effect has been achieved. 

The measure of accuracy was the number of correct answers per task for each participant. Participants' answers to the study questions were transformed into a binary representation with 0 indicating a wrong and 1 a correct option. Answers' binary representation for the RQ1's questions (multiple selections were allowed) consisted of as many binary digits as the available options for participants to select, excluding the ``none'' option, while for the rest of questions' types consisted of a single digit. Participants' performance in each question was computed as the number of occurrences of digit 1 in their response. 

Participants' response time was measured (in seconds) from the moment the visualisation was displayed until the final answer was selected. For each question, participants also rated their confidence on a 1-5 scale with increasing level of confidence (1:not at all, 2:slightly, 3:somewhat, 4:fairly, 5:completely). We remapped this to a $-2$ - 2 scale to center the parameterization.

\subsubsection{Bayesian Analysis}
We conducted a Bayesian analysis of the collected data (the analysis code and data can be found in \cite{taka2022}), which was split into two sub-sets based on the condition (IG and SG). The analysis was conducted on the level of the individual tasks. Fig.~\ref{fig:analysis_models} presents the graphs of the three probabilistic models used for the analysis. More details about the models used for the analysis are provided in the supplemental material.

The accuracy observations were binary values and the propensity of a participant to give a correct answer to each of the tasks was estimated. Each groups' performance in each task was modelled by a binomial likelihood. The posterior {\itshape probability of success}\footnote{This probability expresses the probability of a participant to identify correctly the existence or not of a relation between two variables, or the type of relation, or specific structural information.} $\theta$ of the binomial likelihood was estimated for each group. The two groups were compared in terms of accuracy by taking the differences of the $\theta$'s posterior distribution of each group.

The response time observations were times (in sec). Each groups' response time in each task was modelled by a normal likelihood. The posterior distribution of {\itshape effect size (Cohen's d)} was estimated for the comparison of the two groups to normalise for the varying duration (and thus typical variances) of the tasks. 

The confidence observations were ordinal values. Each groups' response time in each task was modelled by a normal likelihood. Note that we made the simplifying assumption that the ordinal values could be treated as if they lay on a common continuous scale; hence the normal likelihood. A more sophisticated analysis could have inferred a (potentially per-subject) monotonic relationship between ordinal responses and ``true'' confidence. The posterior {\itshape mean confidence level} was estimated for each group as confidence takes ordinal values and there was no need to normalise. The two groups were compared in terms of confidence by taking the differences of the mean confidence posterior distribution of each group. 

Comparing the two groups based on the differences of the posterior distributions, an effect of interaction is more likely given the data as the value $0.0$ becomes less likely under the posterior.

\begin{table*}[!t]
    \centering
    \caption{Summary of probabilistic models and tasks used in the user study. The models' definitions and graphs are presented in the first two columns and the task id, research question each task addresses, and question asked in the rest columns in the order presented to participants.}
    \label{tab:questions}
 \begin{tabular}{p{4.5cm}cccp{5.7cm}}
    \hline
     Problem & Graph & Task & RQ & Question\\
    \hline
    \multirow{5}{*}{
    \makecell{\textbf{Problem 1} \\ 
    \\
    \(\operatorname{a} \sim \operatorname{Uniform}(\operatorname{lower} = 80,\) \\ 
    \(\operatorname{upper} = 100)\) \\
    \(\operatorname{b} \sim \operatorname{Normal}(\mu = 2, \sigma = 10)\) \\
    \(\operatorname{c} \sim \operatorname{Half-Normal}(\sigma = 10)\) \\
    \(\operatorname{temperature} \sim \operatorname{Normal}(\mu = \operatorname{b}\),\\ \(\sigma = \operatorname{c})\)
    }} & \multirow{5}{*}{\includegraphics[width=0.17\textwidth]{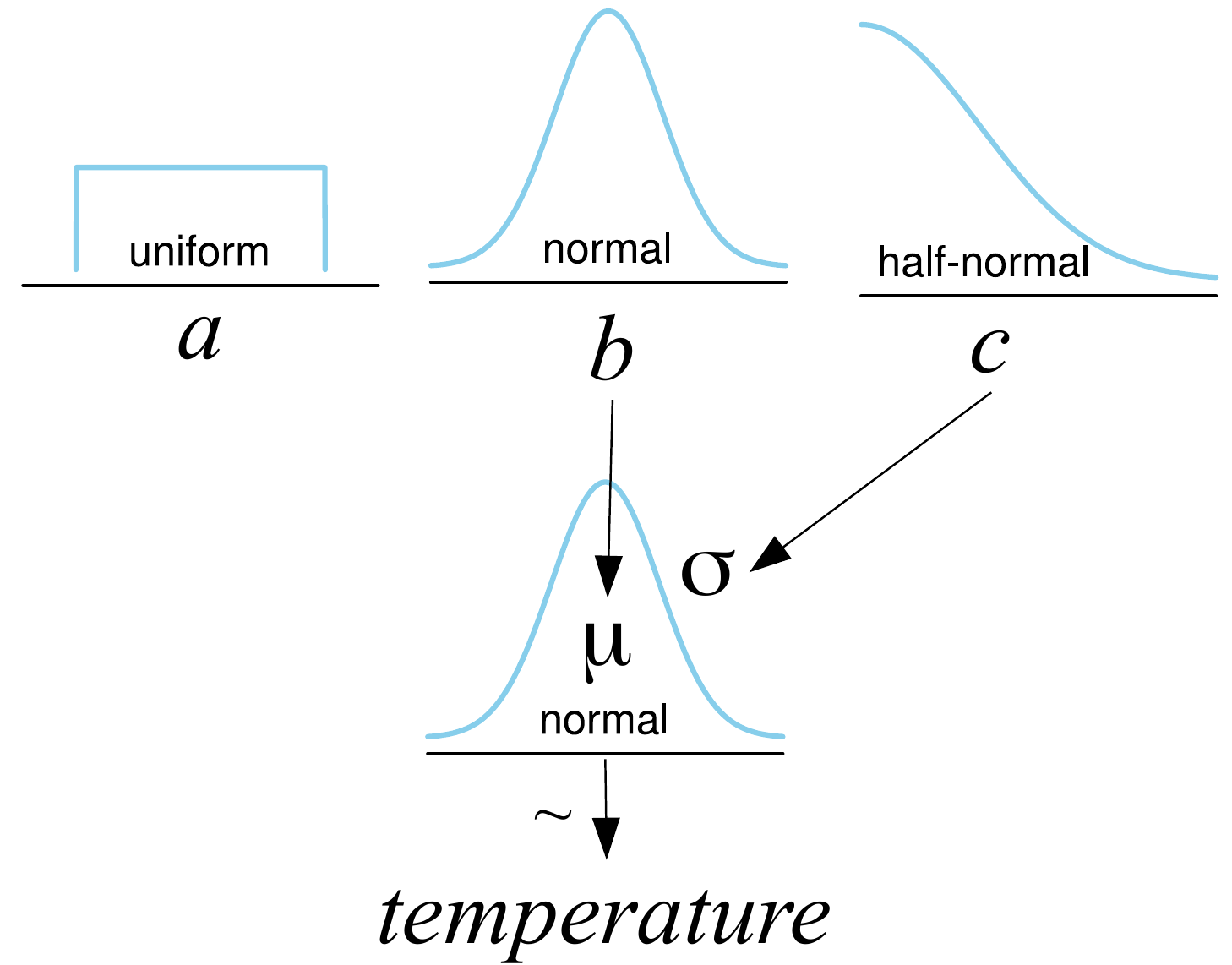}} & \\ && t1 \cellcolor{Gray}& RQ1 \cellcolor{Gray}& Which of the parameters \(\operatorname{a}\), \(\operatorname{b}\) and \(\operatorname{c}\) are related to \(\operatorname{temperature}\)?\cellcolor{Gray}\\ 
    && t2 & RQ2 & How is parameter \(\operatorname{a}\) related to \(\operatorname{temperature}\)?\\
    && t3 \cellcolor{Gray}& RQ2 \cellcolor{Gray}& How is parameter \(\operatorname{b}\) related to \(\operatorname{temperature}\)?\cellcolor{Gray}\\ 
    && t4 & RQ2 & How is parameter \(\operatorname{c}\) related to \(\operatorname{temperature}\)?\\ 
    && t5 \cellcolor{Gray}& RQ3 \cellcolor{Gray}& How would you describe the effect of parameters \(\operatorname{a}\), \(\operatorname{b}\) and \(\operatorname{c}\) on \(\operatorname{temperature}\)?\cellcolor{Gray}\\ 
    \hline
    \multirow{6}{*}{
    \makecell{\textbf{Problem 2}\\
    \\
    \(\operatorname{a} \sim \operatorname{Normal}(\mu = 0, \sigma = 10)\) \\
    \(\operatorname{b} \sim \operatorname{Half-Normal}(\sigma = 10)\) \\
    \(\operatorname{c} \sim \operatorname{Half-Normal}(\sigma = 20)\) \\
    \(\operatorname{random\_number} \sim \operatorname{Uniform}(\) \\ \(\operatorname{lower} = \operatorname{a} - \operatorname{c},\) \\
    \(\operatorname{upper} = \operatorname{a} + \operatorname{c})\)}} &
    \multirow{6}{*}{\includegraphics[width=0.2\textwidth]{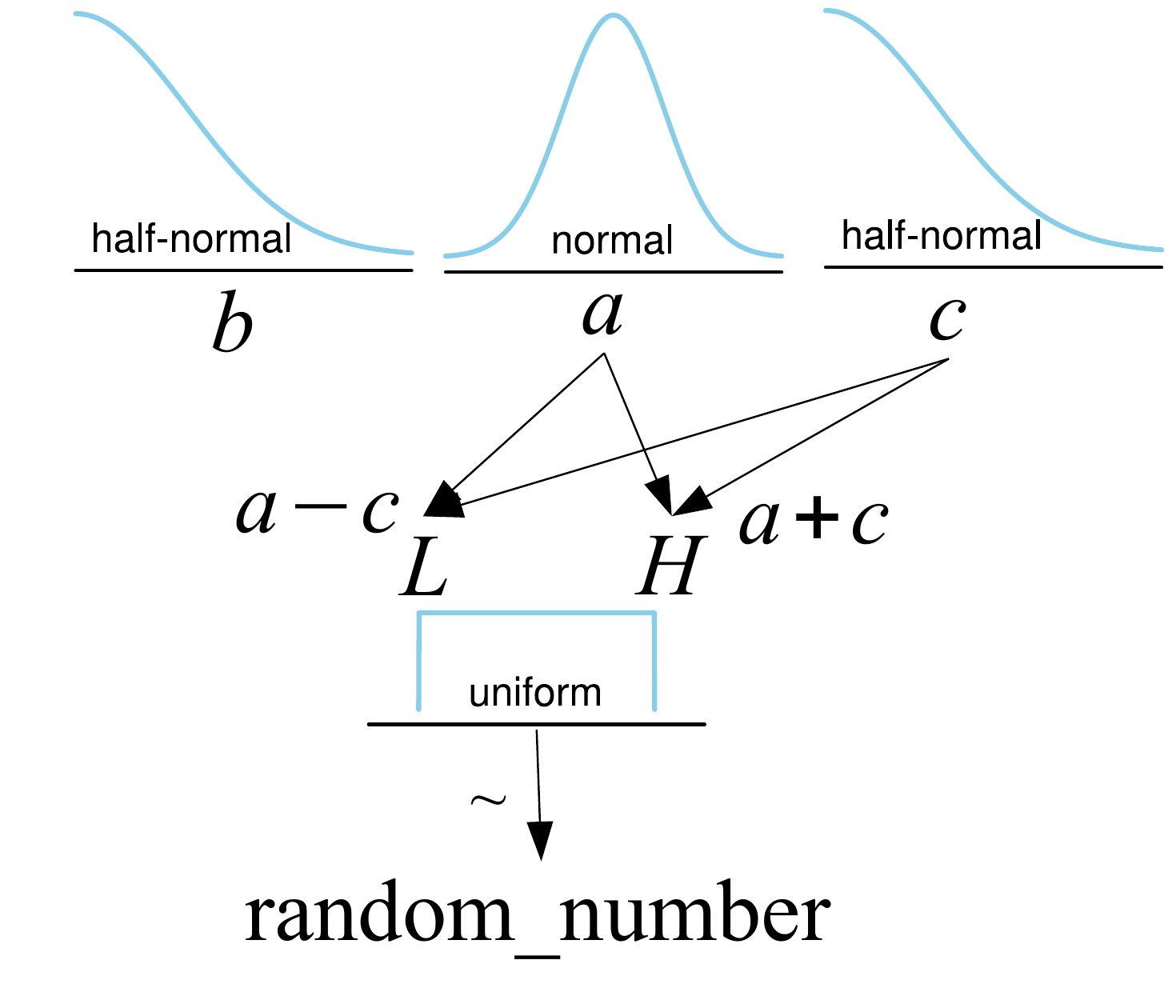}} & t6 & RQ1 & Which of the parameters \(\operatorname{a}\), \(\operatorname{b}\) and \(\operatorname{c}\) are related to \(\operatorname{random\_number}\)?\\ 
    && t7 \cellcolor{Gray}& RQ2 \cellcolor{Gray}& How is parameter \(\operatorname{a}\) related to \(\operatorname{random\_number}\)?\cellcolor{Gray}\\ 
    && t8 & RQ2 & How is parameter \(\operatorname{b}\) related to \(\operatorname{random\_number}\)?\\ 
    && t9 \cellcolor{Gray}& RQ2 \cellcolor{Gray}& How is parameter \(\operatorname{c}\) related to \(\operatorname{random\_number}\)?\cellcolor{Gray}\\ 
    && t10 & RQ3 & How would you describe the effect of parameters \(\operatorname{a}\), \(\operatorname{b}\) and \(\operatorname{c}\) on \(\operatorname{lower\_bound}\)?\\ 
    && t11 \cellcolor{Gray}& RQ3 \cellcolor{Gray}& How would you describe the effect of parameters \(\operatorname{a}\), \(\operatorname{b}\) and \(\operatorname{c}\) on \(\operatorname{upper\_bound}\)?\cellcolor{Gray}\\
    \hline
    \multirow{8}{*}{
    \makecell{\textbf{Problem 3} \\
    \\
    \(\operatorname{c} \sim \operatorname{Normal}(\mu=100, \sigma=150)\) \\
    \(\operatorname{e} \sim \operatorname{Half-Normal}(\sigma=150)\) \\
    \(\operatorname{f} \sim \operatorname{Normal}(\mu=10, \sigma=100)\) \\
    \(\operatorname{g} \sim \operatorname{Half-Normal}(\sigma=100)\) \\
    \(\operatorname{h} \sim \operatorname{Half-Normal}(\sigma=200)\) \\
    \(\operatorname{a_i} \sim \operatorname{Normal}(\mu=\operatorname{c}, \sigma=\operatorname{e})\) \\
    \(\operatorname{b_i} \sim \operatorname{Normal}(\mu=\operatorname{f}, \sigma=\operatorname{g})\) \\
    \(\operatorname{sigma_i} \sim \operatorname{Half-Normal}(\sigma=\operatorname{h})\) \\
    \(\operatorname{d} \sim \operatorname{Normal}(\mu = 0, \sigma = 10)\) \\
    \(\operatorname{reaction\_time_i} \sim \operatorname{Normal}(\) \\
    \(\mu=\operatorname{a_i} + day \cdot \operatorname{b_i},\) \\
    \(\sigma=\operatorname{sigma_i})\)}} & \multirow{8}{*}{\includegraphics[width=0.2\textwidth]{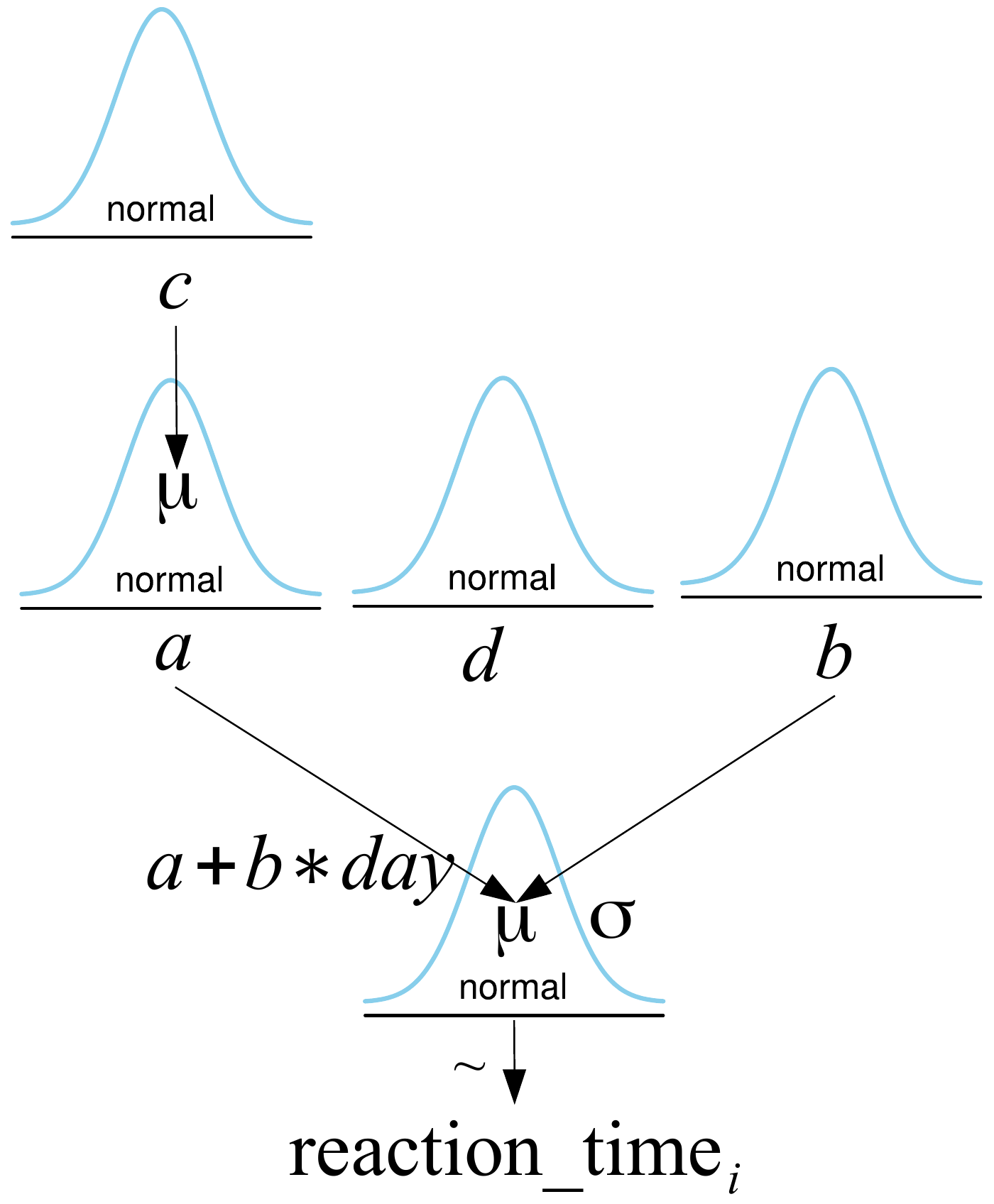}} & t12 & RQ1 & Which of the parameters \(\operatorname{a}\), \(\operatorname{b}\), \(\operatorname{c}\) and \(\operatorname{d}\) are related to \(\operatorname{reaction\_time}\)?\\ 
    && t13 \cellcolor{Gray} & RQ1 \cellcolor{Gray} & Which of the parameters \(\operatorname{b}\), \(\operatorname{c}\) and \(\operatorname{d}\) are related to \(\operatorname{a}\)?\cellcolor{Gray} \\ 
    && t14 & RQ2 & How is parameter \(\operatorname{a}\) related to \(\operatorname{reaction\_time}\)?\\ 
    && t15 \cellcolor{Gray}& RQ2 \cellcolor{Gray}& How is parameter \(\operatorname{b}\) related to \(\operatorname{reaction\_time}\)?\cellcolor{Gray}\\ 
    && t16 & RQ2 & How is parameter \(\operatorname{c}\) related to \(\operatorname{reaction\_time}\)?\\ 
    && t17 \cellcolor{Gray}& RQ2 \cellcolor{Gray}& How is parameter \(\operatorname{d}\) related to \(\operatorname{reaction\_time}\)?\cellcolor{Gray}\\ 
    && t18 & RQ3 & If \(\operatorname{reaction\_time}\), \(\operatorname{a}\) and \(\operatorname{c}\) lie on a graph, what is the structure of the graph?\\ 
    && t19 \cellcolor{Gray}& RQ3 \cellcolor{Gray}& How would you describe the effect of parameters \(\operatorname{a}\), \(\operatorname{b}\) and \(\operatorname{day}\) on \(\operatorname{reaction\_time}\)?\cellcolor{Gray}\\ \\ \\
  \hline
\end{tabular}
\end{table*}

\begin{figure*}[!t]
  \centering
  \subfloat[]{%
        \includegraphics[width=0.58\textwidth]{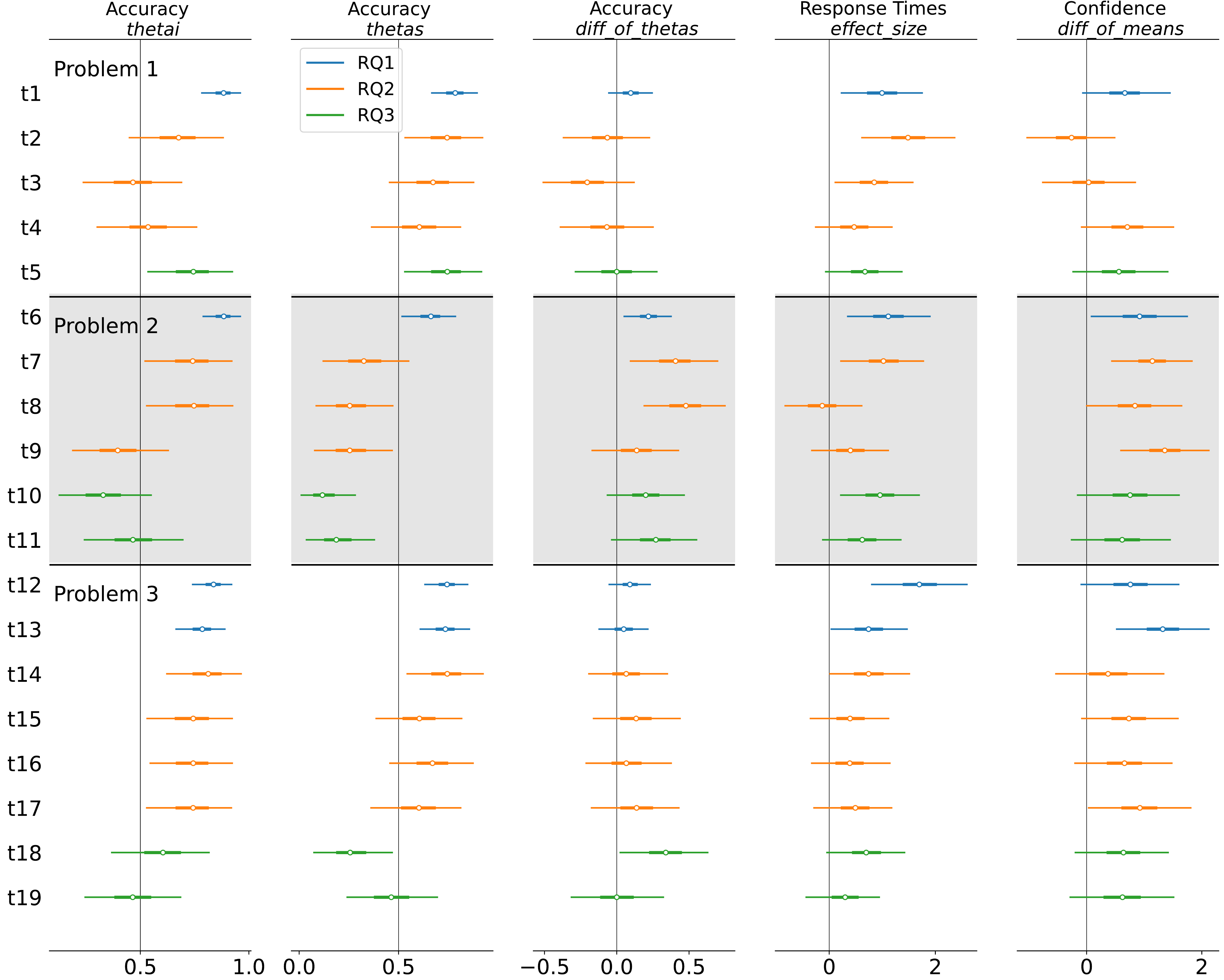}}%
   \subfloat[]{%
        \includegraphics[width=0.42\textwidth]{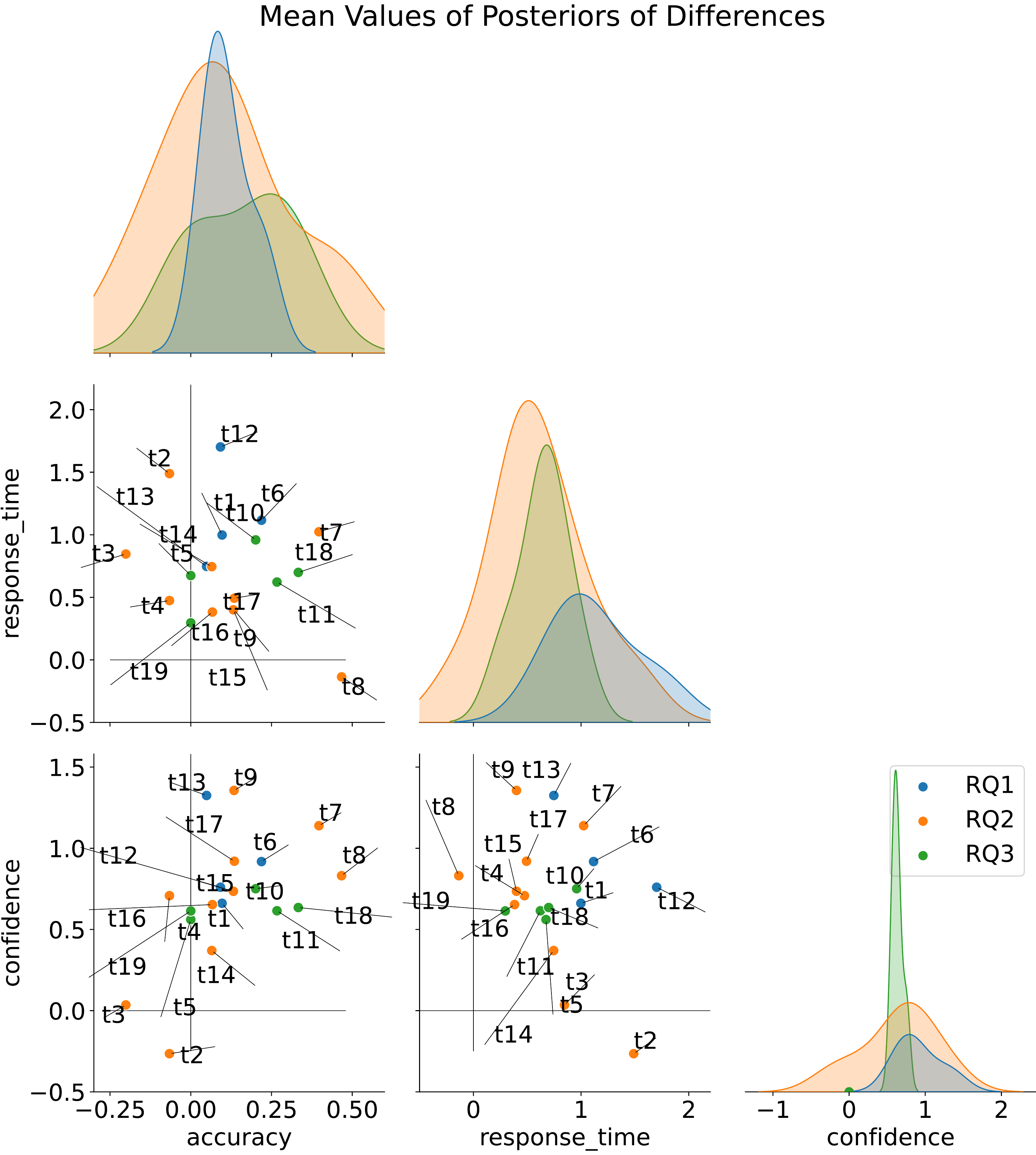}}%
  \caption{(a)Forest plot ($94\%$ highest density intervals) of the posterior distributions of the probability of correct answer for IG ($thetai$) and SG ($thetas$), difference of $theta$s ($diff\_of\_thetas$), $thetai - thetas$, effect size of response times ($effect\_size$) between IG and SG (normalised difference of duration), and difference of the estimated mean confidence of participants about their responses ($diff\_of\_means$). Tasks are presented vertically grouped per problem. (b) Pair plot of mean values of the posteriors of $diff\_of\_thetas$ for the accuracy, $effect\_size$ for the response times and $diff\_of\_means$ for the confidence.
  }
  \label{fig:inference}
\end{figure*}

\subsubsection{Results of Accuracy Analysis}
Based on the accuracy-related forest plots in Fig.~\ref{fig:inference}a, participants estimated performance in overall is good in both groups with the estimated probability $theta$ of giving a correct answer being over $0.5$ in most tasks. An exception to this is the tasks of Problem 2, where both groups do not seem to perform well, but with the IG group seeming to perform better than the SG. Another exception is the last two tasks of Problem 3. Both these cases concern more complicated instances of statistical modeling, than the more trivial cases of statistical associations (e.g. setting the average value or standard deviation of the likelihood) present in the rest of the tasks. Problem 2 was using a parameterization for setting the bounds of a Uniform likelihood and Problem 3 a hierarchical structure.

Observing the differences of the $theta$s in Fig.~\ref{fig:inference}a, it seems there is an obvious effect of interaction in tasks of Problem 2. In some tasks of this problem the effect is stronger (``t6'', ``t7'', ``t8'') and in others weaker (``t9'', ``t10'', ``t11''). Interaction seems to have a strong effect in question ``t18'' of Problem 3. This question expected participants to infer the hierarchical structure between a hyper-prior and prior of the model. 

Tasks ``t2'' (Problem 1), ``t8'' (Problem 2) and ``t17'' (Problem 3) expected participants to identify the absence of relation between the unrelated parameters and the observed variables of the models. The effect of interaction for ``t8'' seems strong, but not for the rest two tasks. The conic shape of the pair plot of the unrelated parameter and observed variable in task ``t8'' (see corresponding figure in supplemental material for task ``t8'' and similar example in Fig.~\ref{fig:prob_model_2}b) might have misleadingly make participants in SG to infer the existence of relation, while the use of interaction by the participants in IG helped into the identification of the absence of relation.

\subsubsection{Results of Response Times Analysis}
Based on the response times-related forest plot in Fig.~\ref{fig:inference}a, participants in the IG seem to need considerably more time to infer lower level of structural detail in comparison to those in the SG. As the level of structural detail increases, the differences of the two groups seem to be pooled towards the reference value. This might imply that in cases of more complex models and structures, the use of interaction would not necessarily bring longer response times. 

\subsubsection{Results of Confidence Analysis}
Based on the confidence-related forest plot in Fig.~\ref{fig:inference}a, interaction seems to have an effect on participants' confidence of response in overall with those in IG being more confident than those in SG. The differences in confidence between the two groups generally seem to be pooled towards the reference value as the level of structural detail increases and we move towards tasks of RQ3. 

A strong effect of interaction on participants' confidence in the lower level of structural detail tasks of Problem 2 (``t6'', ``t7'', ``t8'', ``t9'') seems to exist. There is also a strong effect of interaction in task ``t13'' of Problem 3, although this time there is no corresponding effect in regards with accuracy. This task concerned the relation between a hyper-prior and prior of the Problem 3 model. Although participants in both groups have similar performance in this task, interaction seems to make those using interaction more confident.

\subsubsection{Comparative Analysis of Accuracy, Response Times and Confidence}
An important aspect of the analysis is the investigation of relations between the response time and accuracy or confidence and between the accuracy and confidence. Do higher response times imply better accuracy or higher confidence? Does higher confidence imply better accuracy and vice versa? The conduction of a causal analysis of these parameters is out of the scope of this study, but we will investigate the existence of relations (correlations) between these pairs. This will be done by looking at the correlations of the inferred data.

Fig.~\ref{fig:inference}b presents the pair plot of the mean values of the posteriors of differences for the accuracy, response times, and confidence  between the two groups. Based on the scatter plot of \(\operatorname{response\_time}\) and \(\operatorname{accuracy}\), we could say that any increase in the accuracy of the IG would not be attributed to increased response times in any level of structural detail. 

Similarly, based on the scatter plot of \(\operatorname{response\_time}\) and \(\operatorname{confidence}\), we could say that any increase in the confidence of the IG would not be attributed to increased response times in any level of structural detail. The scatter plot of \(\operatorname{accuracy}\) and \(\operatorname{confidence}\) would imply a slight tendency of increased confidence with increased accuracy of the IG in comparison to the SG especially in RQ2 tasks. This might imply that the increase in participants' confidence in the IG might be partly attributed to the increase in their accuracy, and not solely to the use of interaction.

\subsubsection{Analysis of Interaction Logs}
We conducted an analysis (\cite{taka2022}) of the interaction logs of the IG, which were tracking the coordinates of the selection boxes drawn by the IG participants in each task. Participants in the IG generally were using the selection boxes drawing tool with the (Q1,Q2,Q3) quartiles of the number of selection boxes drawn per task being (4.5, 9., 13.) and of the normalized length of selection boxes\footnote{Lengths of selection boxes were normalized by the range of the corresponding variable.} being (0.11, 0.16, 0.24). No further valuable conclusion could be drawn by this analysis.

\subsection{Limitations of Study}

The user study was designed to include a variety of probabilistic models' types (parameterized, linear regression, hierarchical), distributions (normal, half-normal, uniform), and statistical and mathematical associations (setting the mean, standard deviation, or bounds of the likelihood directly or through simple mathematical equations). A different distribution was used for the unrelated variables in each problem. There are many more model types (logistic regression, GPs), distributions (discrete distributions like binomial and Poisson) and configurations that could be explored in the context of a study like the one presented in this paper. We had to limit the number of questions to ensure the completion of study by participants in roughly an hour.

We limited ourselves to visualisations of the prior distributions in our experiments, to more clearly identify structural relations. Supporting posterior exploration would have different challenges.

Our choice of the type of distributions was limited by the fact that prior sampling from heavy tail distributions (student-t, Pareto, Cauchy) was giving a Dirac delta looking estimation of the probability density. Exploring such options in the prior space and in an interactive framework like the one used by this user study would be pointless, as users would not be able to observe any effect on the distribution of these variables while they would interact.   

IPP does not have any inherent mechanism of exploiting any structural information from the model's graph to arrange variables on the visualization grid in a structure-relevant way like IPME does. The lack of this implicit structure-related visual information might have increased the difficulty of the tasks and made participants feel less confident about their responses.  

The participants' sample of this user study present limited demographics in respect with the age and educational background. We cannot be sure what the results of this study would look like if the sample was more diverse. 

\section{Discussion}{\label{sec:discussion}}
The analysis of the participants' accuracy in their responses suggests that the effect of interaction could become stronger as the model or structures become more sophisticated. The effect of interaction in tasks of Problem 2 seems plausible and strong in the cases of inferring lower level of structural details. This problem was using a parameterization for setting the bounds of a Uniform likelihood, which participants were more unlikely to be familiar with. Most of the tasks in the rest of problems concerned more trivial statistical associations (e.g. setting the average value or standard deviation of the likelihood) which participants could be more familiar with. 

The results also suggest that interaction can considerably improve the performance of users in identifying hierarchical relations in comparison to users who use static visualizations. In the cases of unrelated variables, the effect of interaction seems to be dependent on the form of their prior distribution. Participants in the IG performed considerably better in identifying an unrelated half-normally distributed parameter in comparison to those in the SG, than a uniformly or normally distributed unrelated parameters. The reason for this could be that the shape of the pair plot of a uniformly or normally distributed unrelated parameter and the observed variable would more easily reveal the absence of relation in the static condition. This would not be so explicit in cases of more unusual shapes like the conic one of the half-normally distributed unrelated parameter in Problem 2.

The analysis of the participants' confidence in their responses suggests that the effect of interaction on users' confidence is overly strong by improving their confidence especially in tasks of inferring lower level of structural detail and in tasks of more sophisticated designs like in Problem 2. An interesting finding of the analysis of confidence was that there was a case where participants in the two groups performed similarly, but the participants in the interactive condition had noticeably more confidence about their responses. The analysis of the relations between the inferred differences for the accuracy and confidence between the two groups suggests that there might be a relation between these two parameters implying that the increase in users' confidence in the interaction group might be partly attributed to the increase in their accuracy.

The analysis of the response times suggests that interaction does not necessarily require considerably more time to respond to tasks for inferring higher levels of structural detail about a probabilistic model. However, users who use interaction need noticeably more time to infer lower level of structural details than those in the static condition. Based on the analysis of the relations between the inferred response times and accuracy or confidence, longer response times do not seem to suggest higher accuracy or confidence of users about their responses. This provides an extra piece of evidence that the improved accuracy or higher confidence for users in the interactive condition could be attributed to the element of interaction and not the fact that users were spending more time to explore and comprehend the structure in question.

The interaction logs' analysis showed that the IG participants generally were using the selection box drawing tool. The recorded interaction data could not provide us with more insight into the ways this was used. For example, we do not know if and to what extent IG participants  were combining information from both the pair plots and marginal distributions, or if they were changing their answer or confidence while they were interacting.

We believe that interactive visualizations could and will play a significant role in the field of probabilistic modeling evoking the need for more research to understand how users can be benefited from them. A variety of interactive primitives, model designs, experimental designs that make use of conditional questions repertoires (\cite{cole1989,sedlmeier2001,tsai2011,breslav2014,mosca2021}), the effect of observations in inferring structural information from the posterior, the effect of the strength of variables' relations, the effect of users' statistical background are only few of the parameters that could be investigated to evaluate the benefits of interactive visualizations in this context. Tools like Mimic \cite{breslav2014} for visual analysis of micro-interactions could be used in future studies to provide insight into the ways users read and understand these visualizations. Given the experimental design in this paper, further experimentation could be conducted on a more expanded sample with broader demographics to explore the effect of interaction on users' comprehension of probabilistic models in the broader audience (as Ottley et al. \cite{ottley2012} did for the experimental methodology of Brase \cite{brase2009} and Micallef et al. \cite{micallef2012}). 

In overall, the findings of the analysis provide evidence about the value of interaction in the comprehension of probabilistic models' structure. Interactive visualizations could consist valuable supporting tools in probabilistic modeling and Bayesian analysis making them more accessible to a broader audience. Thus, we believe that this research topic would worth any future research efforts.

\section{Conclusions}{\label{sec:conclusions}}

Interactive tools to support Bayesian analyses are increasingly important both to support analysts' workflow and to communicate results to a wider audience. This has many facets, from communication of uncertainty, representation of high-dimensional posteriors and representation of model structure. We developed the Interactive Pair Plot (IPP) to simultaneously represent the conditional relationships among distributions computed via sample-based Bayesian inference. Our results indicate that interactive visualizations like the IPP can enhance users' comprehension of probabilistic models' structure. The analysis of the user study we conducted indicate that the use of interaction enhances users' comprehension in cases of more sophisticated designs, which are more unlikely users to be familiar with. In particular, interaction helps users identify hierarchical relations among variables and identify unrelated variables, when these are a priori distributed in an unusual way more accurately. Although users using interaction need more time to infer lower level of structural detail than those using a static visualisation, the difference in response times between the two groups seems to become less important as the level of structural detail increases. Users in the interactive condition are more confident about their responses in overall with the effect being stronger in the cases of inferring lower level of structural detail. The findings of this user study provide evidence for the value of interaction in users' comprehension of probabilistic models' structure and pave the way for future investigation into the role of interactivity to support user engagement with Bayesian probabilistic models.

\ifCLASSOPTIONcompsoc
  \section*{Acknowledgments}
\else
  \section*{Acknowledgment}
\fi

This work was supported by the Closed-Loop Data Science for Complex, Computationally- and Data-Intensive Analytics, EPSRC Project: EP/R018634/1. All data and the code for the analysis can be found in \cite{taka2022}.

\ifCLASSOPTIONcaptionsoff
  \newpage
\fi



\bibliographystyle{IEEEtran}
\bibliography{IEEEabrv,bare_jrnl_compsoc}
\end{document}